%% file: ms.tex
\documentclass[1p,sort&compress]{elsarticle}
\usepackage{lineno,hyperref}
\usepackage[normalem]{ulem}
\usepackage{amsmath}
\usepackage{amssymb}
\usepackage{dsfont}
\usepackage{graphicx}
\usepackage{caption}
\usepackage{subcaption}
\usepackage{color}
\usepackage{booktabs}
\usepackage[table,xcdraw]{xcolor}

\renewcommand{\vec}[1]{{#1}}

\newcommand{\J}{\mathbf{J}}
\newcommand{\A}{\mathbf{A}}
\journal{Journal of }

\begin{document}
\input{logo}
\begin{frontmatter}

\title{A Numerical Investigation of Matrix-Free Implicit Time-Stepping Methods for Large CFD Simulations}

\author[cs]{Arash Sarshar\corref{cor1}}
\ead{sarshar@vt.edu}

\author[cs]{Paul Tranquilli}
\ead{ptranq@vt.edu}

\author[aoe]{Brent Pickering}
\ead{bpickeri@vt.edu}

\author[aoe]{Andrew McCall}
\ead{mandrew9@vt.edu}

\author[cs]{Adrian Sandu}
\ead{sandu@cs.vt.edu}

\author[aoe]{Christopher J. Roy}
\ead{cjroy@vt.edu}

\cortext[cor1]{Corresponding author}
\address[cs]{Department of Computer Science and Applications, VirginiaTech}
\address[aoe]{Department of Aerospace and Ocean Engineering, VirginiaTech}

\begin{abstract}
This paper is concerned with the development and testing of advanced time-stepping methods suited for the integration of time-accurate, real-world applications of computational fluid dynamics (CFD). The performance of several time discretization methods is studied numerically with regards to computational efficiency, order of accuracy, and stability, as well as the ability to treat effectively stiff problems. We consider matrix-free implementations, a popular approach for time-stepping methods applied to large CFD applications due to its adherence to scalable matrix-vector operations and a small memory footprint. We compare explicit methods with matrix-free implementations of implicit, linearly-implicit, as well as Rosenbrock-Krylov methods.  We show that Rosenbrock-Krylov methods are competitive with existing techniques excelling for a number of problem types and settings.
\end{abstract}


\end{frontmatter}


\section{Introduction}
\label{sec:Introduction}

While flow problems are inherently unsteady, computer flow simulations have traditionally focused mainly on steady-state flow problems because they reduce the computational effort dramatically. Nevertheless, in many practical applications it is important to quantify the impact of unsteady flow phenomena on the forces and moments exerted on a body. These phenomena impact performance characteristics such as the lift and drag of a body, or the dynamic response of a control system. Historically, in aircraft design, these unsteady effects have required additional analyses to mitigate undesirable aeroelastic effects such as wing flutter and undesirable stall characteristics, among other issues \cite{ja:yurkovich2003flutter,ja:leerausch1996flutter, ja:niewoehner2005stall}.  Sometimes the unsteady effects are beneficial, e.g., when using leading edge extensions (LEX) to improve high angle of attack performance \cite{cp:hirato2015lex}. Studies of low Reynolds number unsteady flows have become much more relevant today with the development of micro air vehicles (MAV) \cite{ja:nakata2011mav, ja:wang2011mav, ja:yang2013mav, cp:percin2012mav, ja:sun2002fruitfly, ja:huang1995lowRe,ja:choi2015lowRe}, with typical sizes as small as 15 cm. At these low Reynolds numbers viscous forces dominate the flow characteristics, leading to unsteady viscous effects such as laminar separation and von K\'{a}rm\'{a}n vortices, as commonly demonstrated in the flow over a cylinder \cite{bk:vonkarman1963aerodynamics}.

The use of CFD allows for preliminary analyses of these designs to determine whether any undesirable unsteady effects will be present, before committing to the expensive development and testing of a physical system. However, a major limitation of unsteady flow analysis using CFD is the prohibitive amount of computational time required to simulate a time-accurate solution with the time integration schemes commonly used to solve the Navier-Stokes equations. This coupled set of nonlinear partial differential equations has to be solved iteratively to determine the solution for each time step. A fine mesh resolution typically required to capture the length and time scales of the flow. For making the computations feasible it is necessary to use a time discretization scheme that maximizes convergence speed without prohibitively restricting the time steps due to stability constraints. It is also important that the time integration scheme provides an accurate solution at every time step.

Explicit time integration methods have been used for time-accurate solutions of unsteady flow problems due to their low computational cost per step and moderate memory requirements. For example, in \cite{kennedy_low-storage_2000} a number of embedded high-order explicit Runge-Kutta methods with minimal memory storage have been developed for the compressible Navier-Stokes equations based on van der Houwen's technique \cite{van1972explicit} for stage  memory storage reduction. However, stability constraints restrict the maximum time steps that explicit methods can employ. 

Implicit time-stepping methods have better stability properties than explicit methods, and therefore they can use very large time steps. However, the computational costs per step are also larger. The overall computational efficiency is given by the tradeoff between the computational cost per step and the total number of steps required to carry out the simulation. 

The main cost of implicit methods is associated with solving a large system of nonlinear equations at each step \cite{Bijl_2002_NSimplicit}. Newton type methods for the solution of nonlinear systems are commonly used in the CFD literature  in conjunction with preconditioned Krylov-based solvers for the inherent linear systems \cite{jothiprasad_higher_order_2003}. The popular Jacobian-free Newton Krylov (JFNK) methods employ finite difference approximations of the Jacobian-vector products required by Krylov solvers \cite{Birken_2012_timeDG}. Studies of JFNK methods applied to solve Navier-Stokes equations \cite{Qin_2000_preconditionNS} have shown that error tolerances of Krylov space solvers need to be carefully optimized for performance and accuracy.

There is considerable interest in developing numerical schemes that provide a suitable level of implicitness for time integration of stiff flow problems, such as to allow relatively large time steps while keeping  the cost per time step comparable to that of explicit methods.  
In this paper we study the  efficiency of several different matrix-free, both explicit and implicit, time integration methods applied to computational fluid dynamics problems  of moderate to large dimensions. In addition to standard techniques, we also examine a new class of lightly-implicit time integration schemes, called Rosenbrock-Krylov methods which are particularly well suited to employ approximate Jacobian-vector products.

The remaining part of this paper is structured as follows. Section \ref{sec:Time_Integration_CFD} reviews numerical methods accessible for the time integration of large systems of ordinary differential equations arising from flow problems. The numerical methods investigated here and their implementation are presented in Section \ref{sec:software}. Section \ref{sec:Num_Exp} applies these methods to a number of test problems and studies their effectiveness in terms of their numerical accuracy, stability, and computational efficiency in case of high dimensional problems. Conclusions and future work directions are discussed in Section \ref{sec:Conclusions}.

\section{Numerical time integration for CFD applications}
\label{sec:Time_Integration_CFD}

Consider the autonomous initial value problem: 
\begin{equation}
 	\label{eqn:ode}
	\frac{d\vec{y}}{dt} = \vec{f}(\vec{y}), \quad y(t_0) = y_0, \quad t_0 \le t \le t_F, \quad y(t) \in \mathbb{R}^N,
	\quad f : \mathbb{R}^N  \to \mathbb{R}^N.
\end{equation}
In this paper equation \eqref{eqn:ode} represents the system of ODEs resulting from the spatial semi-discretization of the Navier-Stokes equations for flow problems in the method-of-lines framework. The system is considered autonomous without loss of generality: any system can be written in autonomous form by appending the time variable to the solution vector. With only time derivatives remaining in equation \eqref{eqn:ode}, it is the choice of time-stepping method that determines the stability, accuracy, and efficiency of the numerical solution as the solution is propagated in time. This paper is concerned with the study of high-order implicit time marching schemes and their performance in large CFD applications. 

We next review several important classes of numerical time integration algorithms.

\subsection{Runge-Kutta methods}
\label{sec:RKmethods}
The historically well-known time integration schemes attributed to Runge and Kutta are well-studied \cite{Butcher_1996_history,Enright_1994_RKsurvey} and extensively utilized in flow applications \cite{jorgenson1989explicit,Kennedy_2000_lowStorageNS}. Let $y_n \approx y(t_n)$ be a numerical approximation of the solution of the system \eqref{eqn:ode}. An $s$-stage Runge-Kutta method(advances the numerical solution to the next time step $t_{n+1}=t_{n}+h$ as follows:
\begin{subequations}
\label{eqn:RK}
\begin{align}
\vec{k}_i& = f\left( \vec{y}_n+ h \sum_{j=1}^{s} a_{i,j}\, \vec{k}_j \right ),\quad i=1,\dots,s; \label{eqn:RK_stage} \\
\vec{y}_{n+1}&= \vec{y}_n + h \sum_{j=1}^{s} b_j\, \vec{k}_j.  \label{eqn:RK_y1}
\end{align}
\end{subequations}
The method coefficients 
\[
\vec{a}={[a_{i,j}]}_{1 \leq i,j \leq s}  \quad \vec{b}={[b_i]}_{1 \leq i \leq s} \quad \vec{c}={[c_i]}_{1 \leq i \leq s},
\]
are determined such that the method \eqref{eqn:RK} has the desired accuracy and stability properties \cite[II.1]{Hairer_book_I}.

Explicit Runge-Kutta (ERK) methods are characterized by coefficients $a_{i,j}=0$ for any $j \le i$.  This means that each stage value $k_i$ \eqref{eqn:RK_stage} depends only on previously stage vectors $k_1,\dots,k_{i-1}$. This leads to the convenient result that explicit Runge-Kutta methods need only one ODE right-hand-side function evaluation per stage, and no linear or nonlinear systems of equations are solved in the process. The stability requirements due to CFL conditions limit the step size $h$, and therefore impact the efficiency of the method.

Singly Diagonally Implicit Runge-Kutta methods (SDIRK) \cite[IV.6]{Hairer_book_II} are characterized by coefficients $a_{i,j}=0$ for any $j < i$, and $a_{i,i}= \gamma > 0$ for all stages $i=1,\dots,s$. Solving for the stage vector $\vec{k}_i$ requires the solution of a nonlinear system of equations at each stage
\begin{align}
\vec{F}_i(\vec{k}_i)&= \vec{k}_i - f\left( \xi_i+ h \gamma \vec{k}_i \right )=0 \quad \text{for} \quad  i=1, \dots, s , \label{eqn:Newton} 
\end{align}
which makes the computational cost per step significantly larger than for ERK. However, this also leads to improved stability properties  and the ability to use much larger time steps.
The nonlinear equation \eqref{eqn:Newton} is solved using Newton-type iterations: 
\begin{equation}
\label{eqn:SDIRK_Newton}
\Delta \vec{k}_i^{\{\ell\}} = - {\left(\frac{\partial {F_i}}{\partial k_i} \right)}^{-1} \vec{F_i}\left(k_i^{\{\ell\}}\right), 
\quad 
\vec{k}_i^{\{\ell+1\}} = \vec{k}_i^{\{\ell\}} + \Delta \vec{k}_i^{\{\ell\}} , \quad \ell = 0,1,\dots
\end{equation}
where
\begin{equation}
\label{eqn:Newton_Jac_Matrix}
\frac{\partial F_i}{\partial k_i} =\mathbf{I}_N- h\,\gamma\, \J_n, 
\end{equation}
and $\J_n$ is the Jacobian of the ODE right-hand-side function:
\begin{equation}
\label{eqn:ODE_Jacobian}
\J_n = \left. \frac{\partial f(y)}{\partial y} \right|_{y=y_n}. 
\end{equation}
The fact that $a_{i,i}=\gamma$ for all stages allows re-using the LU decomposition of \eqref{eqn:Newton_Jac_Matrix} in the solution of linear systems appearing in equation \eqref{eqn:SDIRK_Newton} for all stage vectors $i=1,\dots,s$.

\subsection{Rosenbrock methods}
\label{sec:ROS}

Linearly implicit methods avoid the nonlinear systems \eqref{eqn:SDIRK_Newton} and solve only linear systems at each stage. One step of a Rosenbrock (ROS) method  \cite[IV.7]{Hairer_book_II}  reads:
\begin{subequations}
\label{eq:ROS}
\begin{align}
\vec{Y}_i &=  \vec{y}_n+ h \sum_{j=1}^{i-1} a_{i,j} \,\vec{k}_j, \quad
\vec{k_i} = f\Bigl(\vec{Y}_i \Bigr) + h\, \mathbf{J}_n\, \sum_{j=1}^{i} \gamma_{i,j}\, \vec{k}_j, \quad  i=1, \dots, s; \\
\vec{y_{n+1}}&= \vec{y_n} + h \sum_{i=1}^{s} b_j \vec{k}_j. \label{eqn:ROS_y1}
\end{align}
\end{subequations}
Therefore, the stage vectors $k_i$ are found in succession by solving linear systems of the form:
\begin{equation}
\label{eqn:ROS_stage_eq} 
\left( \mathbf{I}_N - h\, \gamma_{i,i}\, \mathbf{J}_n \right)\, \vec{k}_i = f\left( \vec{Y}_i \right ) +  h\, \mathbf{J}_n\, \sum_{j=1}^{i-1} \gamma_{i,j} \vec{k}_j, \qquad   i=1, \dots, s. 
\end{equation}
As with SDIRK methods, choosing $\gamma_{i,i} = \gamma > 0$ for $i=1,\dots,s$ helps reduce the computational costs by allowing to reuse the same LU factorization  \eqref{eqn:Newton_Jac_Matrix}  for all stages.

\subsubsection{Rosenbrock-Wanner methods}
\label{sec:ROW}
Rosenbrock methods simplify the computational effort necessary to solve the stage vector equations by limiting the implicitness to linear terms containing the exact Jacobian right-hand-side products \cite{Lubich_1995_linearlyImplicit}. As a  consequence, the accuracy of the method depends on the availability of the exact Jacobian. In many practical cases an exact Jacobian is difficult to compute, however some approximation of the Jacobian may be available at reasonable computational cost. The Rosenbrock-Wanner (ROW) methods are Rosenbrock schemes that retain the order of accuracy for any matrix $\A_n$ used in place of the exact Jacobian $\J_n$ in \eqref{eq:ROS}. The preservation of accuracy is possible by imposing additional order conditions on the method coefficients \cite{Rang_2005_Wmethods}. Better approximations of the Jacobian $\A_n\approx \J_n$ will ensure better numerical stability. We note that while the formal definition for a ROW method is the same as in equation \eqref{eq:ROS}, the method coefficients are different due to the additional order conditions. 

\subsubsection{Rosenbrock-Krylov methods}
\label{sec:ROK}

The stage vectors in Rosenbrock-type methods are computed by solving the linear system of dimension $N$ in equation \eqref{eqn:ROS_stage_eq}. For large problems the solutions of these linear systems is best obtained via a Krylov-space iterative linear algebra solver such as GMRES \cite{saad2003iterative}.

Instead of using a Krylov-based iterative solver such as GMRES, Rosenbrock-Krylov (ROK) methods developed in \cite{Sandu_2014_ROK} reformulate the method \eqref{eq:ROS} using implicitness only in the Krylov subspace of dimension $M$ constructed using modified Arnoldi iteration \cite{saad2003iterative}
\[
\mathcal{K}_M\big(\J_n, f(y_n)\big) = \textnormal{range}\{\mathbf{V_n} \}, \quad  \mathbf{V_n} \in \mathbb{R}^{N \times M}, \quad \mathbf{V_n}^T\, \mathbf{V_n} = \mathbf{I}_M, \quad \mathbf{V_n}^T\, \J_n\, \mathbf{V_n} = \mathbf{H_n}.
\]
Here $\bf H$ and $\bf V$ are the upper Hessenberg and the orthogonal basis of the Krylov space, respectively, and are results of Arnoldi process. 

A single time step of a ROK method is constructed as follows \cite{Sandu_2014_ROK}: 
\begin{subequations}
\label{eq:ROK}
\begin{align}
  F_i &= f \left( y_n +\displaystyle \sum_{j=1}^{i-1} \alpha_{i,j}k_j\right), \label{eq:ROK_stage_breakdown}\\
 \phi_i &= \mathbf{V_n}^T \, F_i, \\
   \lambda_i &= \left( \mathbf{I}_{M} - h\,\gamma\, \mathbf{H_n}\right)^{-1}\, \left( h\,\phi_i + h\, \mathbf{H_n} \,\displaystyle\sum_{j=1}^{i-1} \gamma_{i,j}\, \lambda_j\right), \label{eqn:ROK_Lambda_system}\\
       k_i &= \mathbf{V_n}\, \lambda_i + h\, (F_i - \mathbf{V_n}\, \phi_i), \label{eqn:ROK_full_space_stage_vectors}\\
y_{n+1} &=  y_n + \displaystyle\sum_{i=1}^s b_i\, k_i.
\end{align}
\end{subequations}
For $M \ll N$ the linear system  \eqref{eqn:ROK_Lambda_system} is easily solvable using direct methods, and the stage vectors in full space can be recovered projecting the reduced space stage values back to full space\cite{Sandu_2014_ROK}. The minimum dimension of the Krylov space is determined by the desired order of the numerical scheme. Readers interested in the derivation of the order conditions for this method may refer to \cite{Sandu_2014_ROK}. In the extreme case $M=0$ the method \eqref{eq:ROK} reduces to an explicit Runge-Kutta method. In practice, however, we need the Krylov space to be large enough to capture some of the dominant eigenvalues of the full Jacobian, corresponding to fast-changing modes of \eqref{eqn:ode}, such as to alleviate the restrictions on step size of explicit methods imposed by the stiffness of the problem. An important question is whether the additional computational cost required by Arnoldi iteration can be compensated by the increases in step size due to the implicit nature of the method. The numerical results in Section \ref{sec:software} provide answers to this question.

\subsection{Matrix-free implementations}
\label{sec:matfree}

As discussed in Sections \ref{sec:RKmethods} and \ref{sec:ROS}, implicit time-stepping methods use the Jacobian matrix \eqref{eqn:ODE_Jacobian} to  solve linear or nonlinear systems of equations at each step. The dimension of the state vector in  \eqref{eqn:ode} may become significantly large when a highly refined spatial discretization is required, whether to capture fine details of flow or when CFD is used in large data-driven applications such as climate research. Computation and storage of Jacobian matrices, even in sparse form, is not favorable in such scenarios. 

In some cases the complexity of the spatial discretization scheme impedes construction of analytic Jacobian matrices. In CFD codes such as SENSEI-Lite \cite{derlaga2013sensei}, the use of complex upwind flux schemes including Roe's and Van Leer's flux scheme \cite[V]{laney1998computational} in addition to MUSCL \cite{ja:vanleer1979limiters}reconstruction with flux limiters make the generation of an analytic Jacobian challenging. Furthermore, to allow for modularity as a research code, SENSEI-Lite allows these different flux schemes and limiters to be used interchangeably, dependent upon the test problem configuration. This interchangeability would necessitate the formulation of an analytic Jacobian for every possible configuration, eliminating the modularity of the code structure. The framework of matrix-free methods allows us to exploit the benefits of advanced time-stepping methods without forming the Jacobian matrix directly.

Krylov space iterative methods for solving the stage equations \eqref{eqn:SDIRK_Newton} or \eqref{eqn:ROS_stage_eq} rely on computing Jacobian-vector products. Instead of computing the Jacobian matrix and then multiplying, it is possible to approximate directly Jacobian-vector products using the finite-difference approximation of a directional derivative: 
\begin{equation}
\label{eqn:matvecFD}
\J_n \cdot v= \left.\frac{\partial f(y)}{\partial y}\right|_{y_n}\cdot v \approx \frac{f(y_n+\varepsilon v)-f(y_n)}{\varepsilon}.
\end{equation}
The optimum $\varepsilon$ is chosen considering the trade-off between truncation and round-off errors \cite{KnollKeyes}. Higher-order approximations are not favorable here as they require more right-hand-side function evaluations, which, considering the large dimensions of the problem, are costly to compute. 

A more in-depth analysis of different strategies  to compute Jacobian-vector products, and their effects on convergence and efficiency of implicit time integration methods, can be found in \cite{Tranquilli2016}. Of special interest is using exact Jacobian-vector products instead of finite difference approximations. Numerical experiments applied to discretizations of PDEs in \cite{Tranquilli2016} indicate that exact Jacobian-vector products provide more robustness in observed convergence orders, and increased runtime efficiency over methods using approximate products \eqref{eqn:matvecFD}.

\section{The software infrastructure for numerical investigations}
\label{sec:software}
%
\subsection{Time-stepping schemes and their implementation}
The time integration software used in the numerical experiments is \textsc{matlode} \cite{matlodeWeb}, a Matlab library for integration of ODE systems including implicit and explicit Runge-Kutta methods, Rosenbrock methods and Krylov based methods. The package also supports forward, adjoint, and tangent linear models, enabling sensitivity analysis applications. Aside from the methods available in \textsc{matlode} package, Matlab's explicit time-stepping scheme based on Dormand and Prince \cite{Dormand80} is also included in tests for comparison. Table \ref{tab:numerical_methods_overview} summarizes the methods used in numerical experiments and their properties.
\begin{table}[htb]
\centering
\caption{Overview of time stepping methods used in numerical experiments. }
\label{tab:numerical_methods_overview}
\begin{tabular}{llccl}
\hline
\rowcolor[HTML]{EFEFEF} 
Method          			& Family               	& Stages & Order & Stability             \\ \hline
ERK           			& Explicit Runge-Kutta 	& 5      & 4     & Conditionally stable  \\
DOPRI5   			& Explicit Runge-Kutta  & 7      & 5     & Conditionally stable  \\
DOPRI853 			& Explicit Runge-Kutta  & 12     & 8     & Conditionally stable  \\
SDIRK           			& Implicit Runge-Kutta 	& 5      & 4     & L-stable              \\
ROS4            			& Rosenbrock           	& 4      & 4     & L-stable              \\
ROW							& Rosenbrock-W        	& 4      & 3     & L-stable              \\
ROK             			& Rosenbrock-Krylov    	& 5      & 4     & Conditionally stable  \\
ODE45           			& Explicit Runge-Kutta 	& 5      & 4     & Conditionally stable  \\ \hline
\end{tabular}
\end{table}
%
\subsection{The fluid flow simulation code}

SENSE-Lite, the CFD code employed in this paper, can solve both the Euler and Navier-Stokes equations for compressible flow \cite{derlaga2013sensei}:
\begin{subequations}
\label{eqn:Nav_Stokes}
\begin{align}
\frac{\partial \rho}{\partial t} +
\frac{\partial \rho v_1}{\partial x_1} + \frac{\partial \rho v_2}{\partial x_2} &= 0 , \\
\frac{\partial \rho v_1}{\partial t} +
\frac{\partial \rho v_1^2}{\partial x_1} + \frac{\partial \rho v_1 v_2}{\partial x_2} + \frac{\partial p}{\partial x_1} &= \frac{\partial \tau_{11}}{\partial x_1} + \frac{\partial \tau_{12}}{\partial x_2} ,\\
\frac{\partial \rho v_2}{\partial t} +
\frac{\partial \rho v_1 v_2}{\partial x_1} + \frac{\partial \rho v_2^2}{\partial x_2} + \frac{\partial p}{\partial x_2} &= \frac{\partial \tau_{12}}{\partial x_1} + \frac{\partial \tau_{22}}{\partial x_2} ,\\
\frac{\partial E_t}{\partial t} +
\frac{\partial \rho v_1 E_t}{\partial x_1} + \frac{\partial \rho v_2 E_t}{\partial x_2} + \frac{\partial v_1 p}{\partial x_1} + \frac{\partial v_2 p}{\partial x_2} &= \notag \\
 k\,\left(\frac{\partial q_1}{\partial x_1} + \frac{\partial q_2}{\partial x_2}\right) + \mu\, \Big(\frac{\partial}{\partial x_1} (v_1 \tau_{11} + v_2 \tau_{12}) &+ \frac{\partial}{\partial x_2} (v_1 \tau_{12} + v_2 \tau_{22}) \Big) ,
\end{align}
where
\begin{align}
q_i &= -k\, \frac{\partial T}{\partial x_i} , \\ \tau_{i,j} &= \mu \,\left(\frac{\partial v_j}{\partial x_i} + \frac{\partial v_i}{\partial x_j}\right) - \frac{2}{3}\, \mu \, \left(\vec{\triangledown} \cdot \vec{v}\right)\, \delta_{i,j}.
\end{align}
\end{subequations}
SENSEI-Lite uses a curvilinear, structured-grid, finite volume method.  Second-order spatial accuracy is achieved through a standard total variation diminishing scheme consisting of MUSCL reconstruction and selectable flux limiters \cite{ja:vanleer1979limiters}. The code is written in C++ and  MEX interfaces are used to call it from within time integrators implemented in Matlab. The primary function of the MEX code is to return the spatial residual for a given solution state; this is returned as a vector that is independent of any temporal information and can be used for building arbitrary time integration methods such as multi-stage ERK. This residual vector can optionally be returned as multiple vectors split according to the underlying equations, e.g., viscous and inviscid contributions from the Navier-Stokes equations as shown in \eqref{eqn:Nav_Stokes}; these vectors sum to the full residual and can also be used to integrate the equations independently. The Matlab client code is responsible for storing multiple solution state and update vectors as necessary, and can apply time-dependent source terms if desired.

\section{Numerical results}
\label{sec:Num_Exp}

\subsection{Experimental setting} 
\label{sec:Setting}

This section details the numerical experiments setup using \textsc{matlode} and SENSEI-Lite packages to study the performance of matrix-free time-stepping methods on unsteady flow problems. A reference solution is computed and stored by integrating the model using an explicit method with tight tolerances ($\sim 10^ {-9}$) on errors. In each numerical experiment, the solution at the final integration time is compared against this reference solution, and the error is measured using the $\mathcal{L}_2$ norm: 
\begin{align*}
Error = \left\Vert{y_N - y_{\rm ref}}\right\Vert_2 \qquad \text{where} \qquad y=\left[ \rho , \rho v , \rho u, E_t \right]^T
\end{align*}

Where $y_N$ and $y_{ref}$ are numerical and reference state vectors at the final integration time respectively. 
All experiments use matrix-free time-stepping methods. For each test problem we evaluate the eigenvalues of the Jacobian matrix in equation \eqref{eqn:ODE_Jacobian} to get information about the dynamic modes of the state variable evolution. A wide spread of the eigenvalues indicates the existence of both slow and fast dynamics, in other words, of ``stiff'' dynamics. The largest eigenvalue gives an estimate of the largest stable step-size in explicit methods.  Eigenvalues are computed using Matlab's implementation of the implicitly restarted Arnoldi method that estimates the largest 1000 eigenvalues in magnitude for each test problem; this computation is also performed in matrix-free form.

\subsection{Vortex-shedding cylinder test problem} 
\label{sec:Cylinder_Intro}

The vortex-shedding cylinder test problem consists of a two-dimensional circular cylinder in a low subsonic flow (Mach 0.1) using a gas model for air at 5,000 ft altitude standard atmospheric conditions (278K, 84.31kPa).  The viscosity for air is calculated based on local flow conditions using Sutherland's law. The modeled is a free-stream flow, using far-field boundary conditions set at over 100 chord lengths away from the surface of the cylinder to minimize interactions with the boundary. The default cylinder diameter is $8 \times 10^{-5}\, m$, which yields a Reynolds number of approximately 200 and results in a steady and predictable two-dimensional shedding of alternating vortices behind the cylinder. At this low Reynolds number there are no sub-grid-scale turbulence effects, so all physically accurate spatial and temporal scales in the solution can be directly modeled; therefore, the CFD code can solve the laminar Navier-Stokes equations with no underlying turbulence model. For all test problems, the flux scheme used is Roe's approximate Riemann solver and no flux limiter is employed in the MUSCL scheme as flux limiters were observed to be unnecessary and to generally reduce the accuracy of the FVM reconstruction in the continuous and smooth flow field around the cylinder. The parameters of this problem are modified to provide different tests, as summarized in Table \ref{tab:cylinder_param}. Care is taken so that the qualitative solution behavior (and appropriate Reynolds number) is maintained for all tests.

{\renewcommand{\arraystretch}{1.2}
\begin{table}[htp]
\centering
\caption{Parameters of the vortex-shedding cylinder test problems.}
\label{tab:cylinder_param}
\begin{tabular}{|l|l|l|l|}
\hline
\multicolumn{1}{|c|}{Parameter} & \multicolumn{1}{c|}{Description} & Test problem 1             & Test problem 2             \\
                                &                                  & \multicolumn{1}{c|}{value} & \multicolumn{1}{c|}{value} \\\hline 
$\rho (kg/m^3)$                 & Reference fluid density          & 1.0565                     & 1.0565                     \\
$S$                             & Sutherland's coefficient         & 1.45E-6                    & 2.9E-6                     \\
$T (K)$                         & Temperature                      & 278                        & 278                        \\
$v (m/s)$                       & Reference velocity               & 340                        & 340                        \\
$L (m)$                         & Diameter                         & 8E-5                       & 8E-5                       \\
$Re$                            & Reynold's number                 & 165.90                     & 82.95                      \\ \hline
\end{tabular}
\end{table}

The first experiment is performed on the vortex-shedding cylinder test problem 1 with parameters given in Table \ref{tab:cylinder_param}. %
Figure \ref{fig:cyl_flow} illustrates snapshots of the density component of the flow for cylinder test problem 1 at different times, showing the cyclic development of vortices behind the cylindrical object. 
%
\begin{figure}
    \centering
\includegraphics[width =\textwidth]{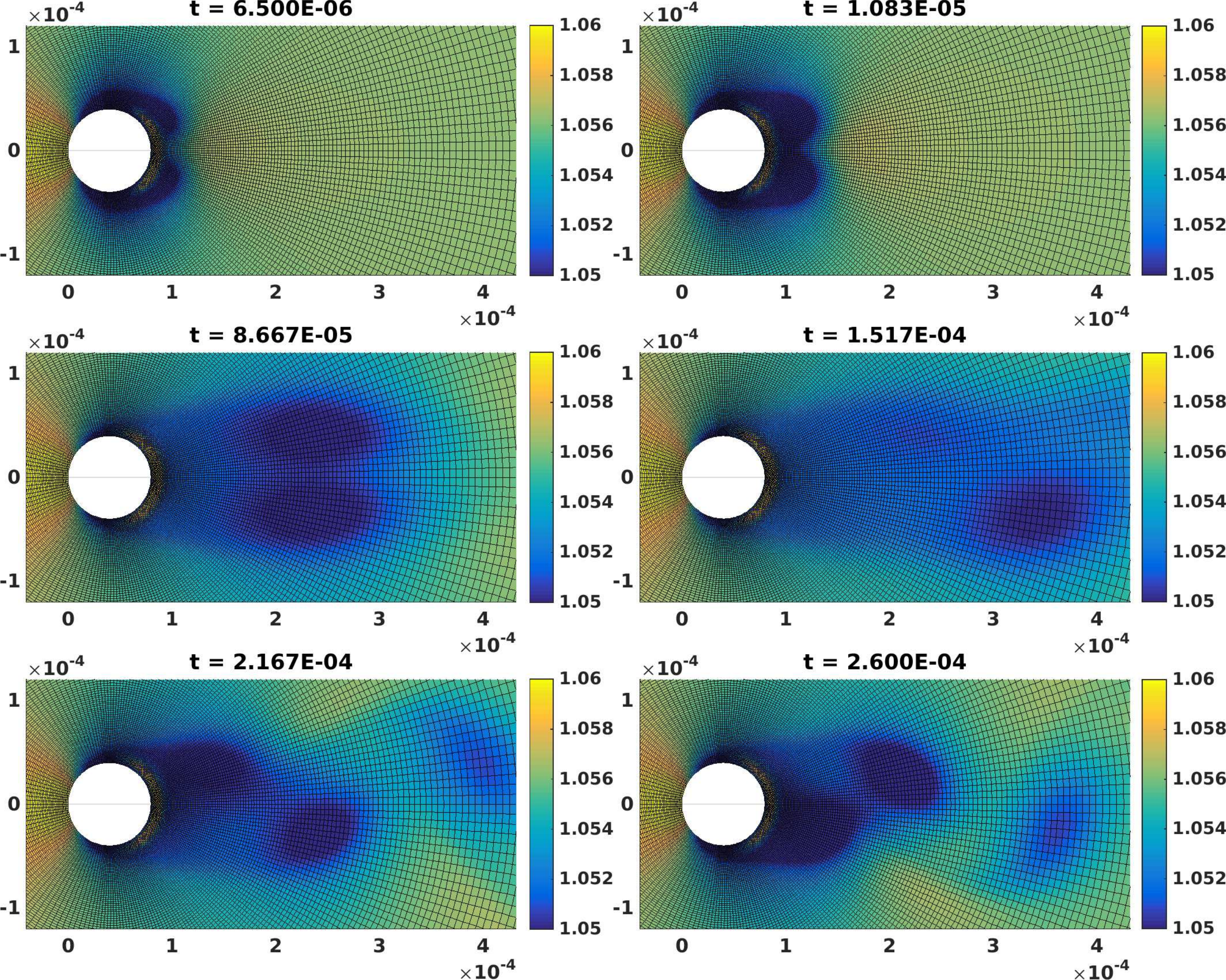}
\vspace*{6px}
\includegraphics[width =\textwidth]{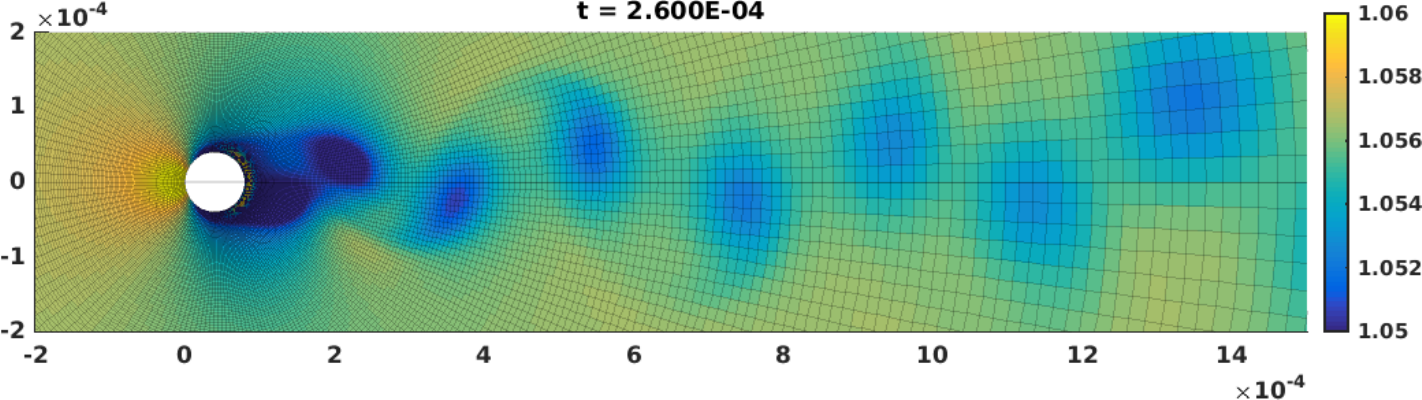}
\caption{Snapshots of density component of the flow for vortex-shedding cylinder test problem 1 at different times.}
\label{fig:cyl_flow}
\end{figure}
%
This experiment uses fixed step sizes to study the temporal orders of convergence for each method, and the results are plotted in Figure \ref{fig:cyl_order}. Numerical orders of convergence calculated for each method are reported in Table \ref{tab:Cyl_conv}. One notable observation is the significantly lower numerical order of convergence for SDIRK method as a result of poor convergence of the Newton iteration, especially in the absence of preconditioners for the solution of linear systems. Readers interested in numerical experiments on a smaller problem that verify the theoretical order of convergence may consult the Appendix.  
%
\begin{figure}
\centering
  \includegraphics[width=0.5 \linewidth]{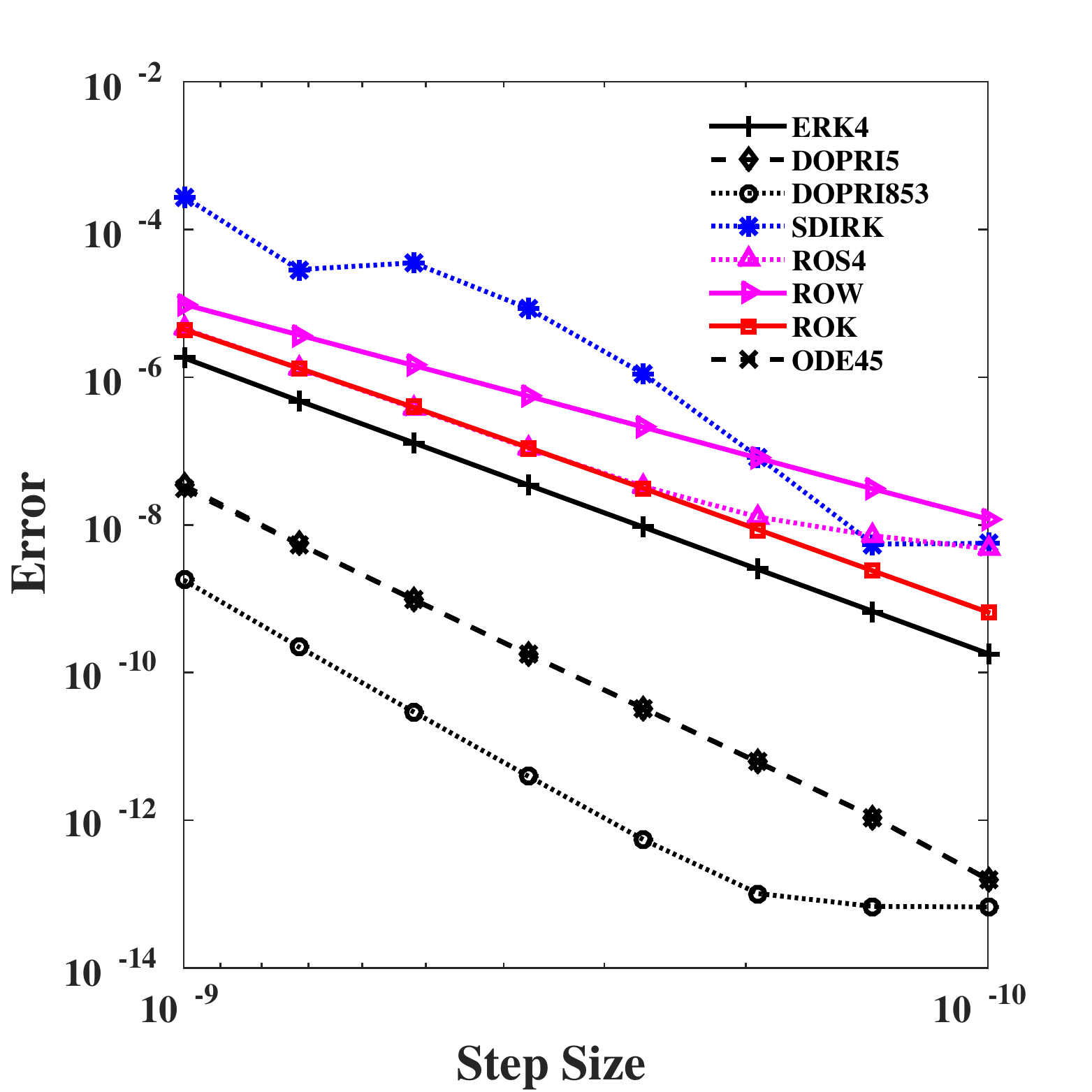}
\caption{Convergence diagrams for the matrix-free methods of Table \ref{tab:numerical_methods_overview} using fixed step sizes over integration window $T=[0,10^{-7}]$ second for cylinder test problem 1.}
\label{fig:cyl_order}
\end{figure}
%
\begin{table}[h]
\centering
\caption{Numerical orders of convergence for various methods applied to the cylinder test problem 1.}
\label{tab:Cyl_conv}
\begin{tabular}{lcc}
\hline
Method          			& Numerical order 	& Theoretical order\\
\hline
ERK4            			& 4.02 			& 4\\
ERK5 (DOPRI5)   			& 5.29			& 5\\
ERK5 (DOPRI853) 			& 5.97 			& 8 \\
SDIRK           			& 2.94			& 4 \\
ROS4            			& 3.11	 		& 4 \\
ROW						    & 2.96 			& 3 \\
ROK             			& 3.85 			& 4 \\
ODE45           			& 5.39 			& 5\\
\hline
\end{tabular}
\end{table}

The cylinder test problem 2 with parameters given in Table \ref{tab:cylinder_param}, uses a different value for Sutherland's Law coefficient that translates into increased kinetic viscosity compared to vortex-shedding cylinder test problem 1. Figure \ref{fig:cyl_eigs} shows the numerical approximation of the first 1000 eigenvalues for the two test problems. As indicated in Figure \ref{fig:cyl_eigs} the more viscous test problem shows larger negative eigenvalues, therefore, we expect stricter stability bounds on the step sizes for this test problem. For both problems the cluster of eigenvalues  with the largest negative real parts consists of only a limited number of modes. This fact becomes relevant when we consider ROK methods that use a reduced order Jacobian for implicit integration. Throughout the numerical experiments the ROK method employs a reduced space of dimension four, unless otherwise specified.
\begin{figure}[htbp]
\centering
\begin{subfigure}{.45\textwidth}
  \centering
  \includegraphics[width=\linewidth]{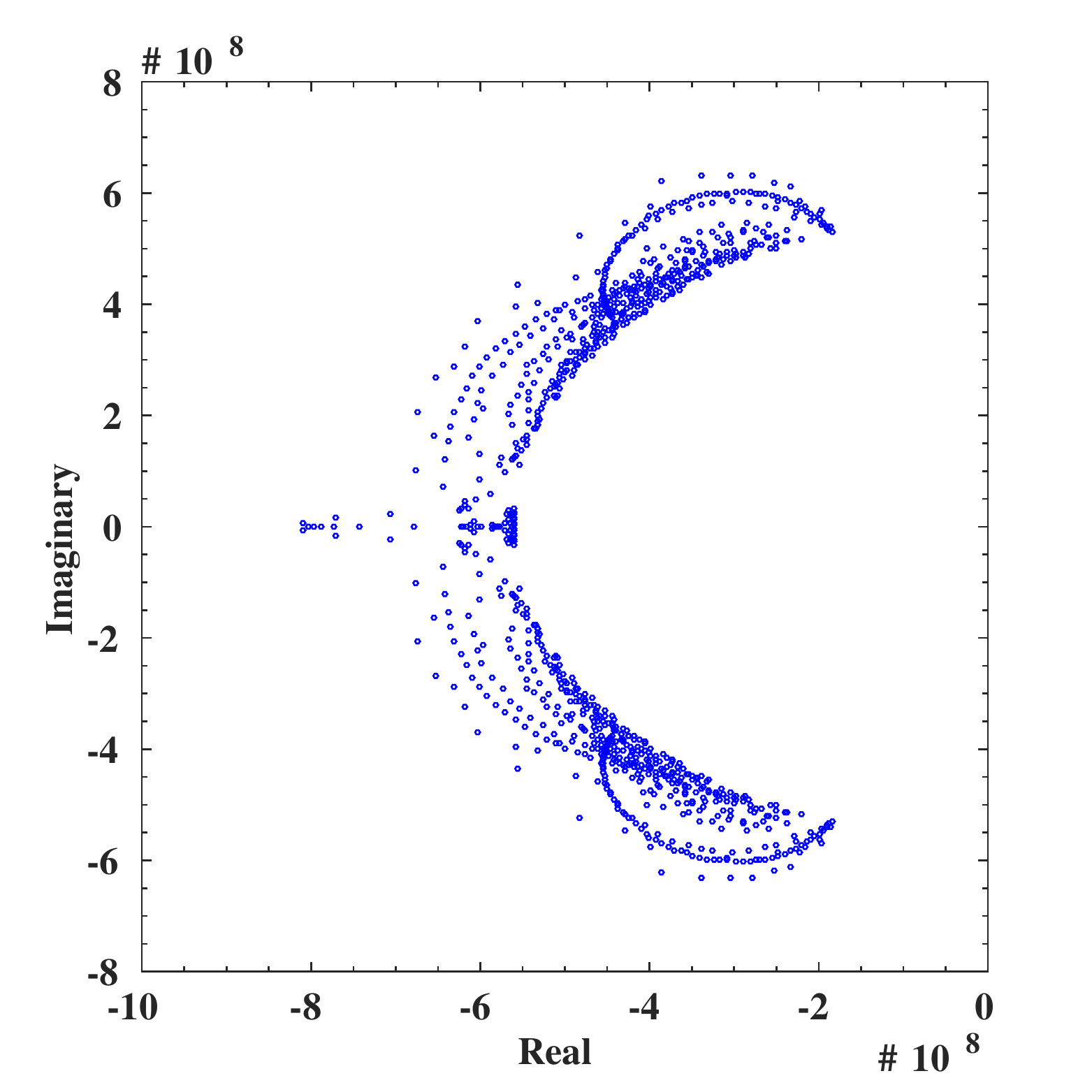}
  \caption{Cylinder test problem 1}
  \label{fig:cyl_eigs_test-1}
\end{subfigure}%
\begin{subfigure}{.45\textwidth}
  \centering
  \includegraphics[width=  \linewidth]{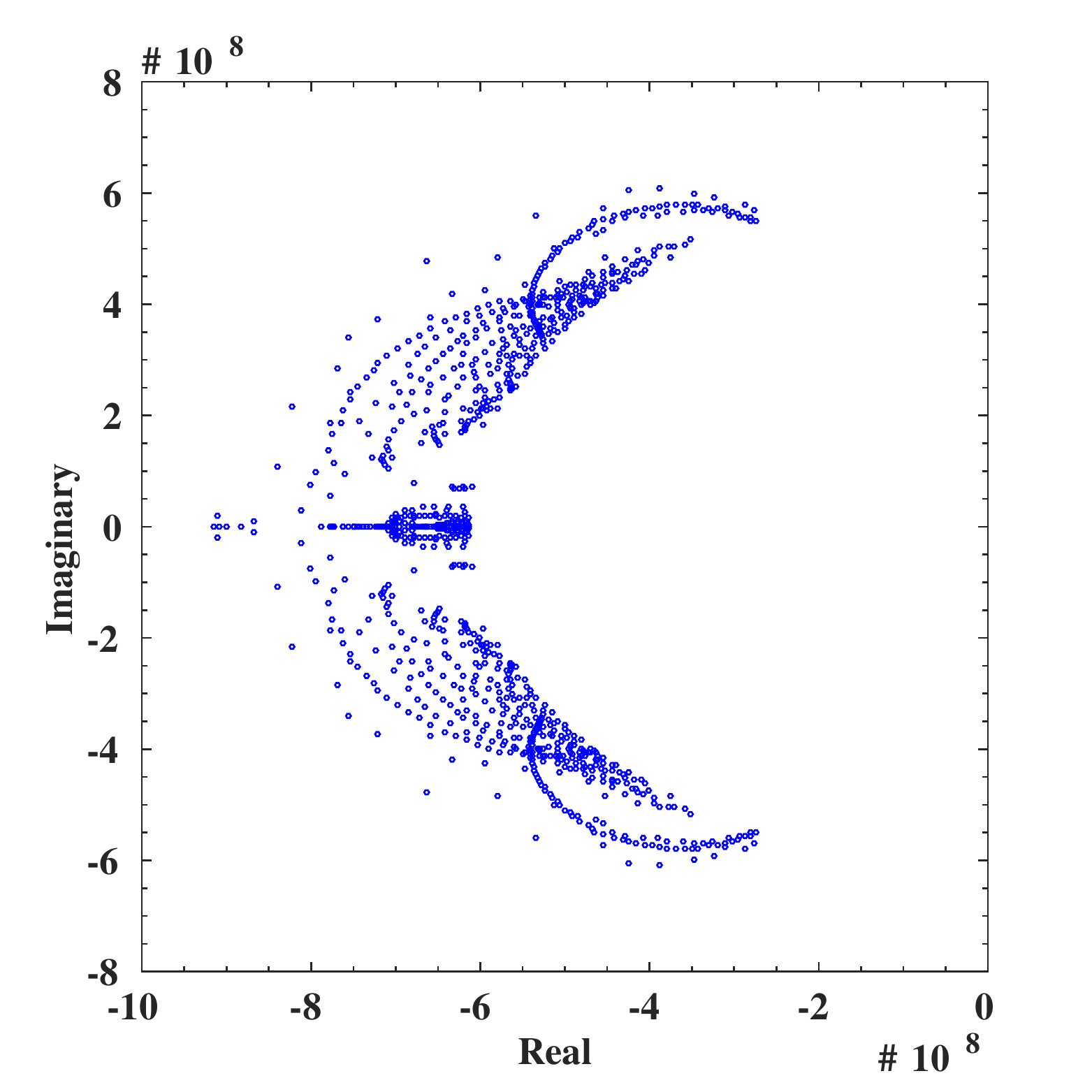}
  \caption{Cylinder test problem 2}
  \label{fig:cyl_eigs_test-2}
\end{subfigure}
\caption{Eigenvalue distribution of the Jacobians of the right-hand-side functions for vortex-shedding cylinder test problems.}
\label{fig:cyl_eigs}
\end{figure}

The stability of a method is related to its choice of timesteps in the adaptive time stepping framework. Figure \ref{fig:cyl_time_steps} compares the step sizes of fourth order explicit Runge-Kutta method (ERK) to Rosenbrock-Krylov method of the same order for two different error tolerances. We can verify that the step sizes for the explicit method quickly reaches the upper bound set by the stability constraints regardless of the accuracy tolerance chosen for the method. On the other hand,  by implicitly treating some of the stiff modes, the ROK method is able to achieve stable numerical integration for larger step sizes as is clear from Figures \ref{fig:cyl_def_ROK_time_steps} and \ref{fig:cyl_2x_suth_ROK_time_steps}. Furthermore, we notice that the step sizes scale well relative to the required accuracy. 
\begin{figure}[htbp]
\centering
\begin{subfigure}{0.45\textwidth}
  \centering
  \includegraphics[width=\linewidth]{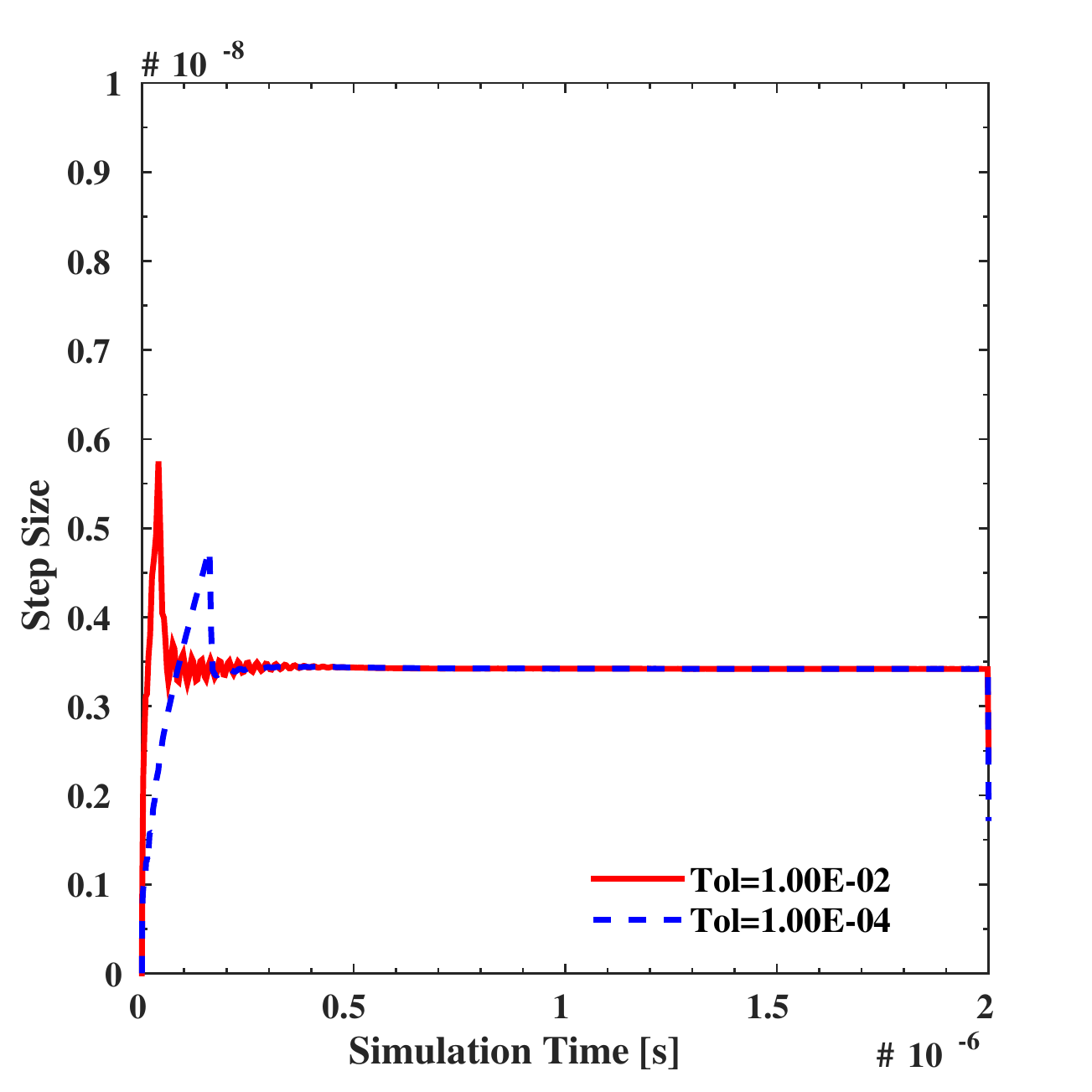}
  \caption{ERK on cylinder test problem 1}
  \label{fig:cyl_def_ERK_time_steps}
\end{subfigure}
~~
\begin{subfigure}{0.45\textwidth}
  \centering
  \includegraphics[width= \linewidth]{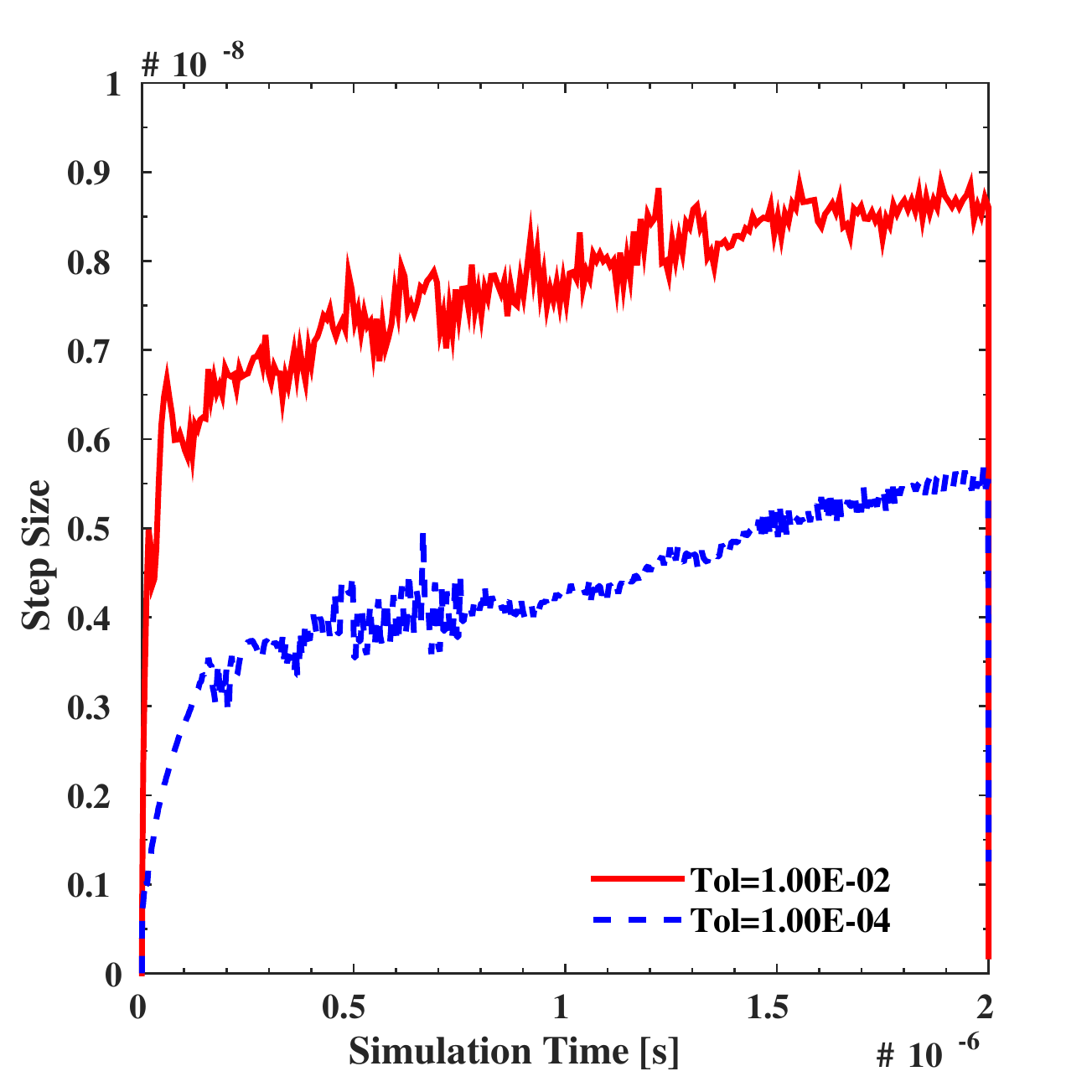}
  \caption{ROK on cylinder test problem 1}
  \label{fig:cyl_def_ROK_time_steps}
\end{subfigure}
\begin{subfigure}{.45\textwidth}
  \centering
  \includegraphics[width=\linewidth]{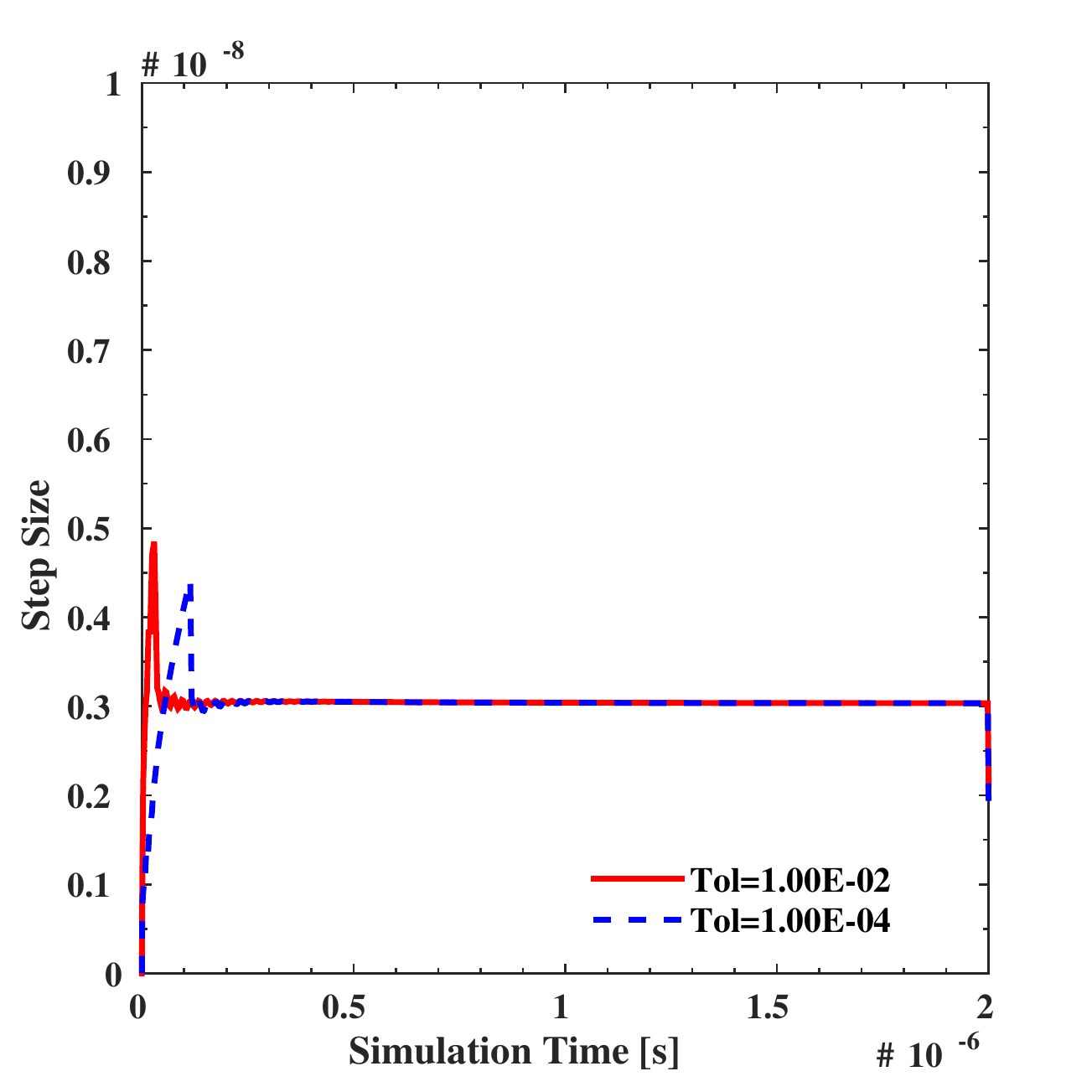}
  \caption{ERK on cylinder test problem 2}
  \label{fig:cyl_2x_suth_ERK_time_steps}
\end{subfigure}
~~
\begin{subfigure}{.45\textwidth}
  \centering
  \includegraphics[width= \linewidth]{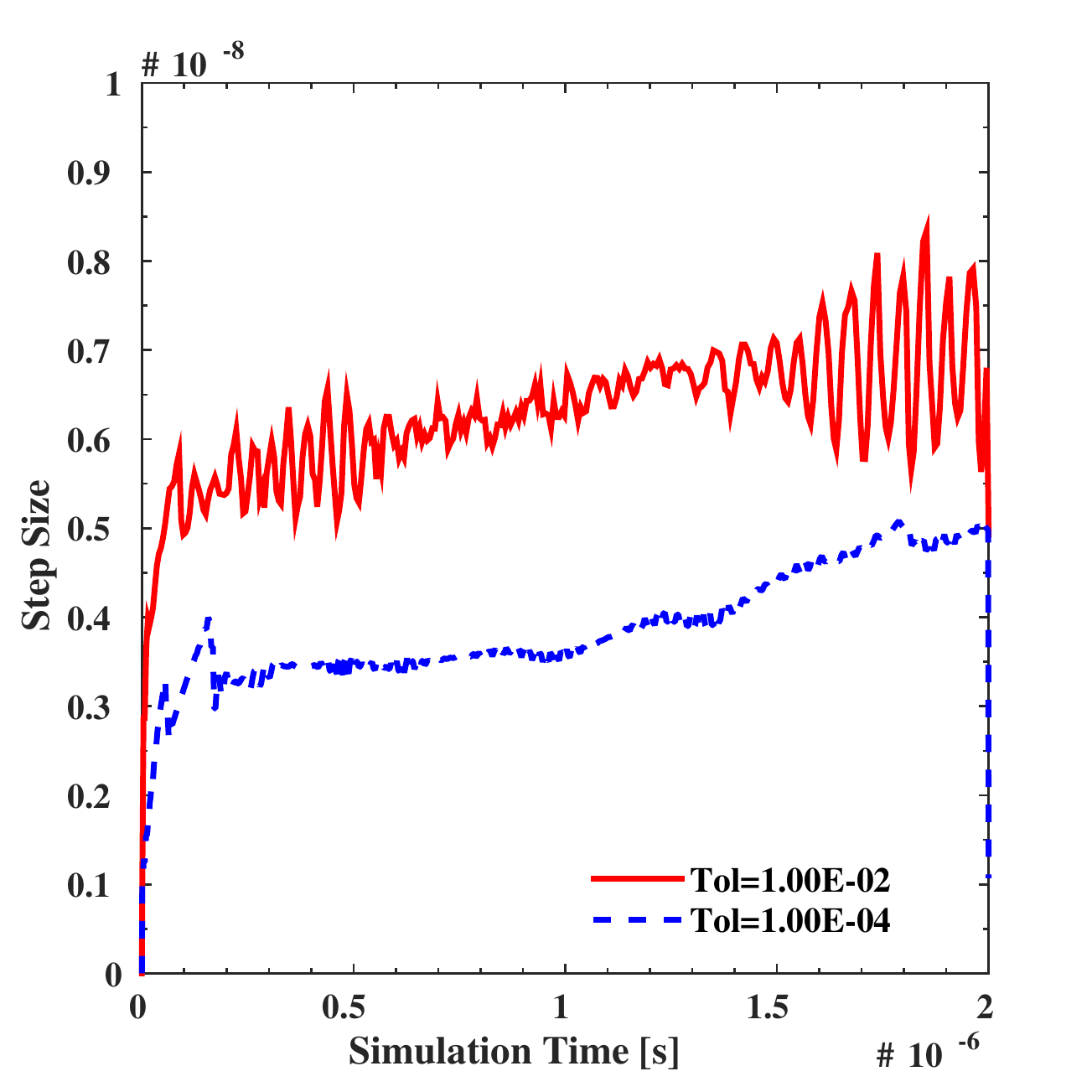}
  \caption{ROK on cylinder test problem 2}
  \label{fig:cyl_2x_suth_ROK_time_steps}
\end{subfigure}
\caption{Adaptive time steps taken by the explicit Runge-Kutta method and by the ROK method for vortex shedding cylinder test problems. }
\label{fig:cyl_time_steps}
\end{figure}

Figure \ref{fig:cyl_M4_steps_timing} shows the Work-precision diagrams for the vortex-shedding cylinder test problems. Integration is performed over time window $T=[0, 2 \times 10^{-6}]$ seconds. These results lead to the following conclusions: 
\begin{itemize}
\item As the adaptive time stepping method uses tighter tolerances, the integrator takes smaller steps leading to an increased number of total steps. This is the case for all integration methods presented here. However, the total number of steps for explicit methods  does not change considerably for a wide range of tolerances, an effect observable in Figure \ref{fig:cyl_M4_steps_timing} where the lines for explicit methods are nearly vertical. This is a result of the fact that for stiff problems the adaptive time steps are bounded by stability requirements rather than by accuracy constraints.    
\item Implicit methods are able to take fewer steps when the required solution tolerances are low, due to their improved stability properties. This is the case for ROS, ROW and SDIRK methods in Figure \ref{fig:cyl_def_M4_steps} and \ref{fig:cyl_2x_suth_M4_steps}.  However, inspecting the runtime diagrams on Figures \ref{fig:cyl_def_M4_timing} and \ref{fig:cyl_2x_suth_M4_timing} reveals that the the increased cost of these methods makes them considerably less efficient.
\item The effect of stiffness of the problem can also be seen in the timing reported in Figures \ref{fig:cyl_def_M4_timing} and \ref{fig:cyl_2x_suth_M4_timing}. We notice that while the explicit methods take about the same amount of time to complete the integration for all choices of solution tolerance, the runtime scales better with tolerances for implicit methods.
\item ROK is the most efficient method for the cylinder test problems for error tolerances below $10^{-6}$.  This is an indication that ROK is able to capture sufficiently many stiff components in its Krylov subspace, and that by treating them implicitly the method is able to take larger time steps.
\end{itemize}

\begin{figure}[htbp]
\centering
\begin{subfigure}{.45\textwidth}
  \centering
  \includegraphics[width=\linewidth]{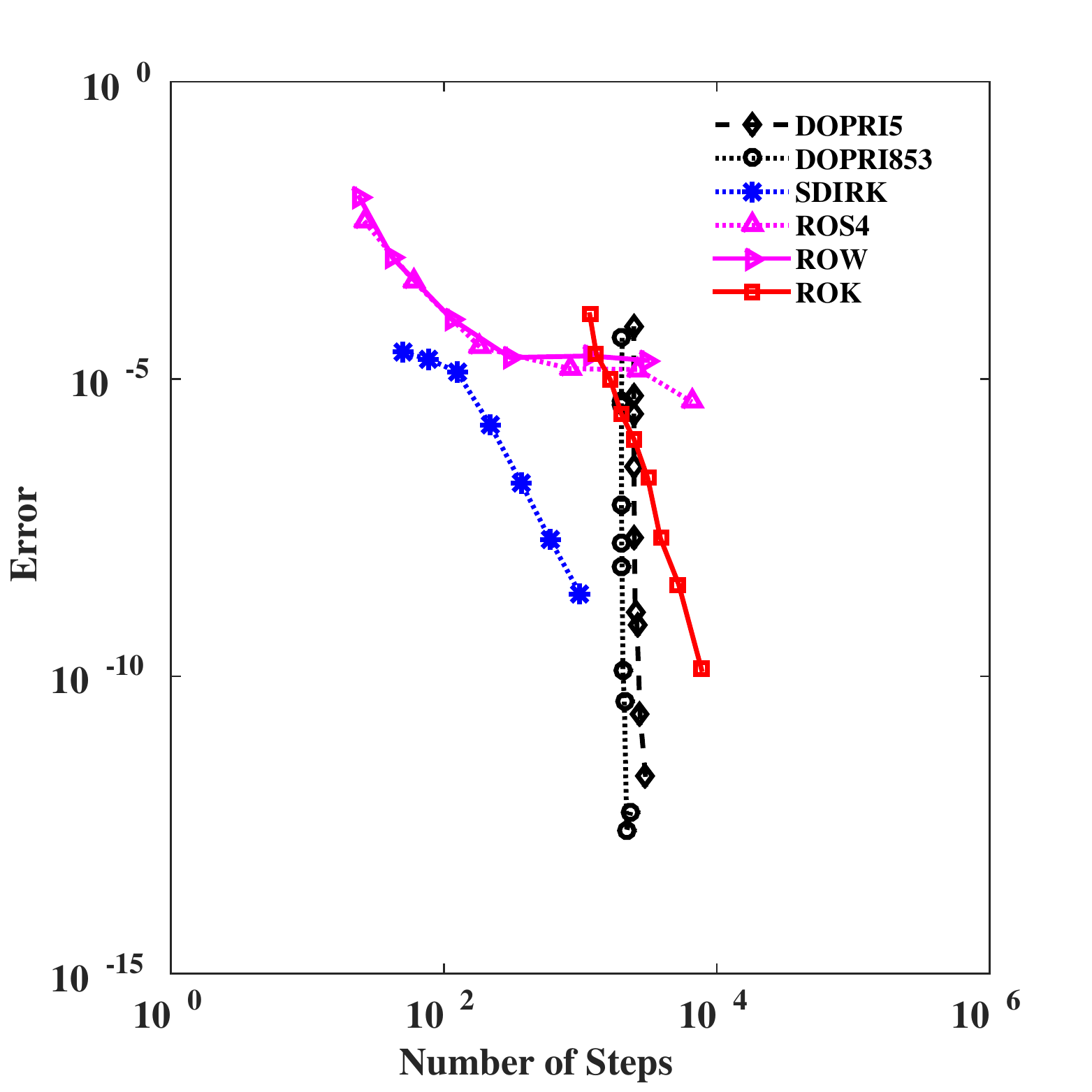}
  \caption{Convergence diagram, cylinder test problem 1}
  \label{fig:cyl_def_M4_steps}
\end{subfigure}
~~
\begin{subfigure}{.45\textwidth}
  \centering
  \includegraphics[width= \linewidth]{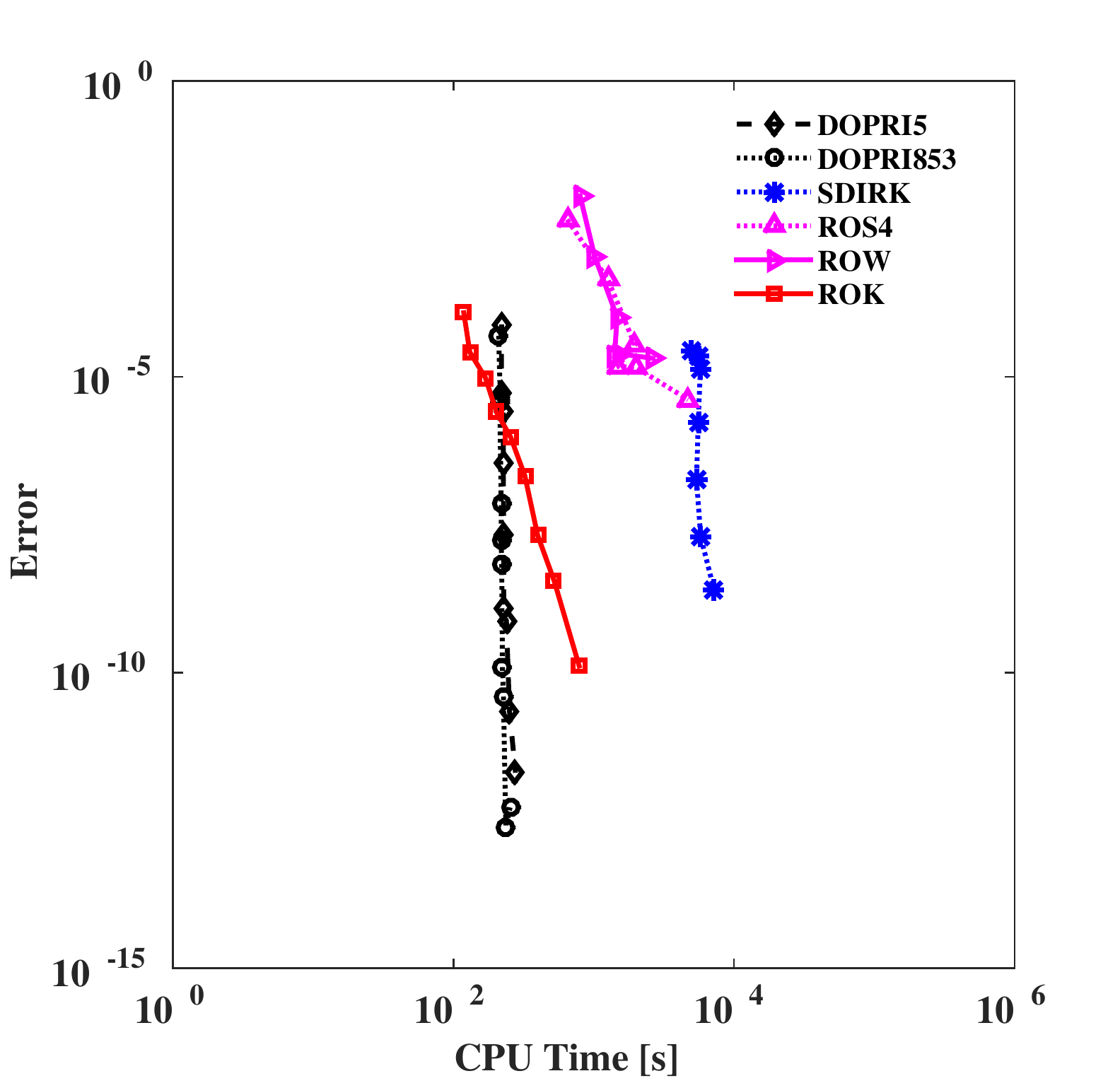}
  \caption{Work-precision diagram, cylinder test problem 1}
  \label{fig:cyl_def_M4_timing}
\end{subfigure}
\begin{subfigure}{.45\textwidth}
  \centering
  \includegraphics[width=\linewidth]{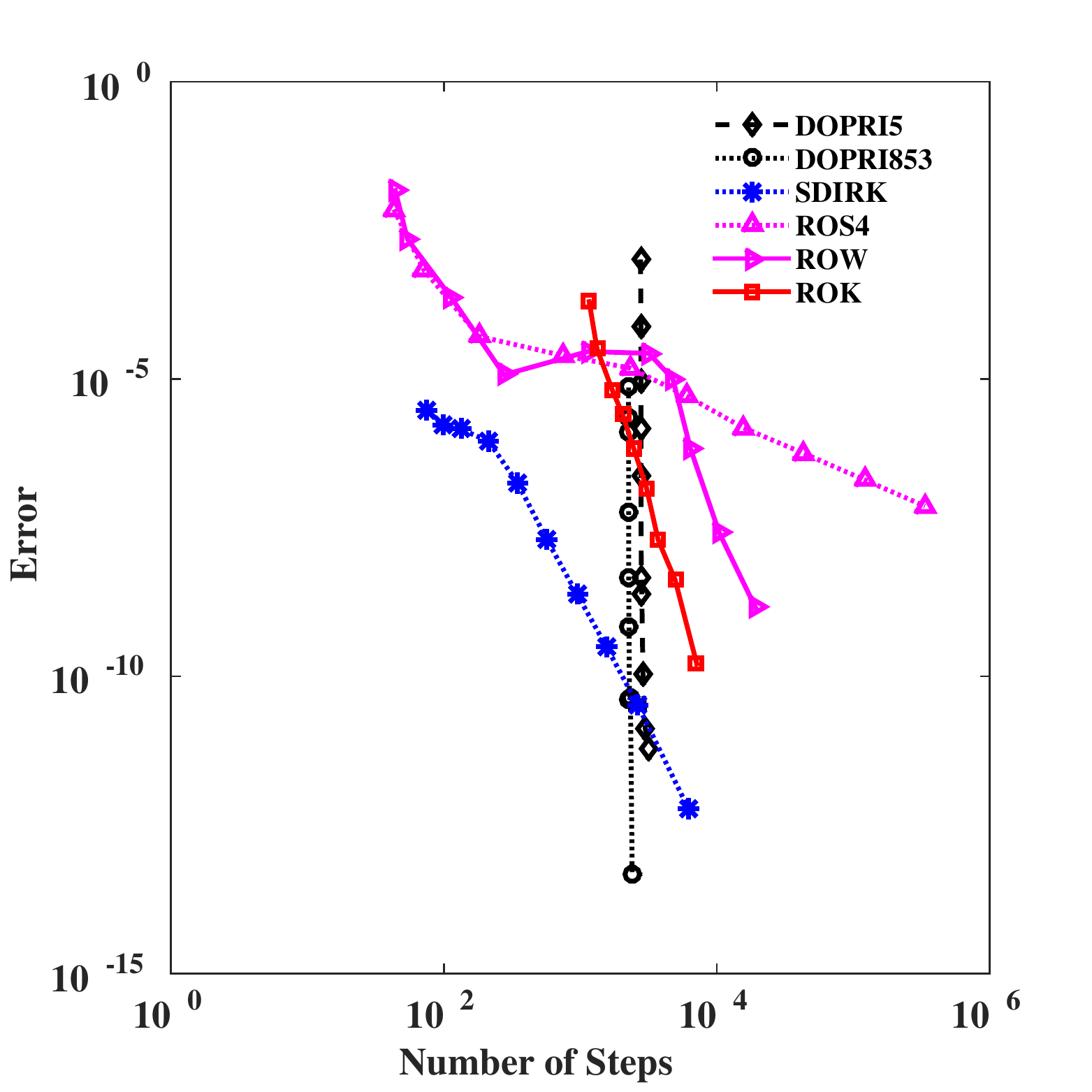}
  \caption{Convergence diagram, cylinder test problem 2}
  \label{fig:cyl_2x_suth_M4_steps}
\end{subfigure}
~~
\begin{subfigure}{.45\textwidth}
  \centering
  \includegraphics[width= \linewidth]{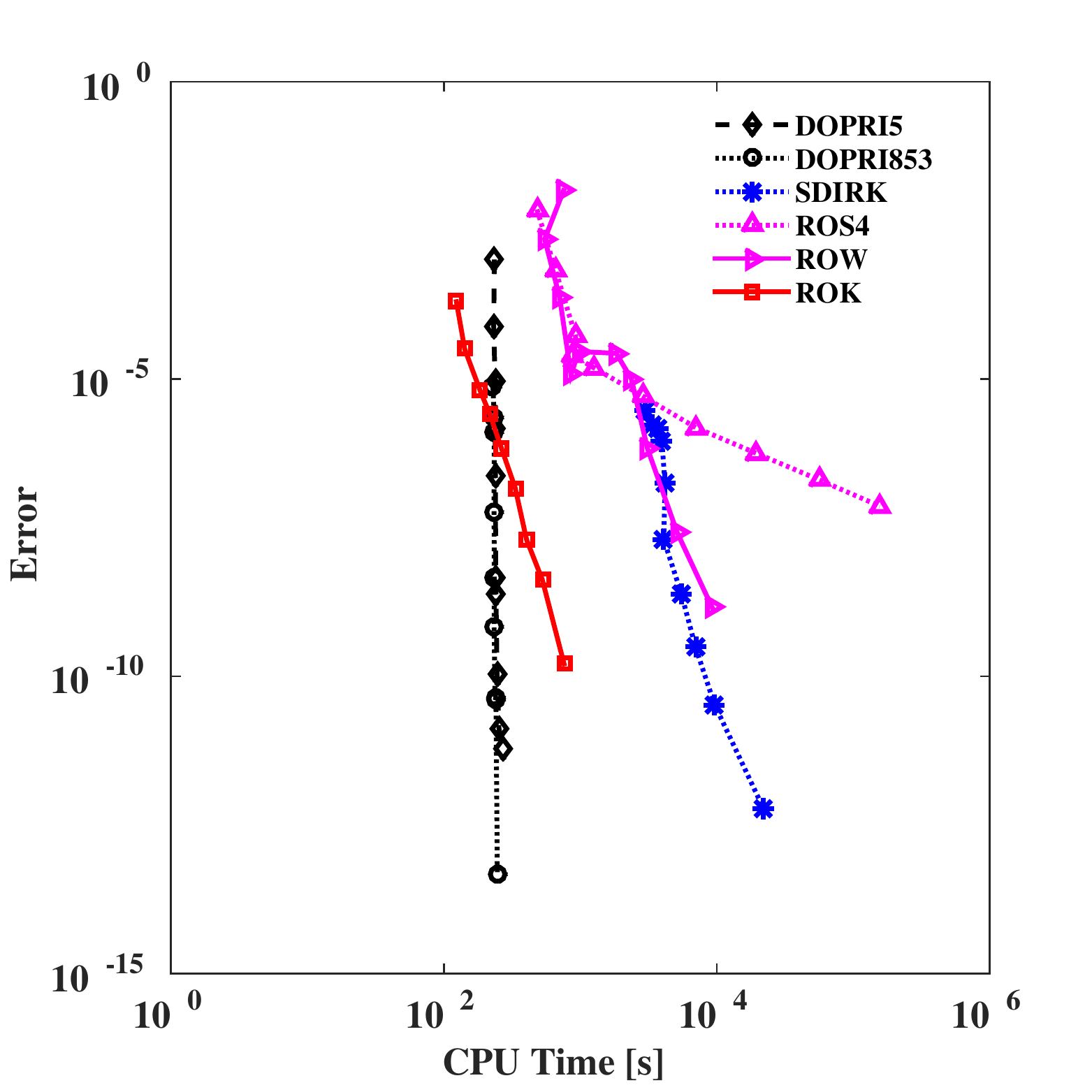}
  \caption{Work-precision diagram, cylinder test problem 2}
  \label{fig:cyl_2x_suth_M4_timing}
\end{subfigure}
\caption{Relative performance of different integration methods applied to the cylinder test problems.}
\label{fig:cyl_M4_steps_timing}
\end{figure}

\subsection{Flow over \textsc{NACA0012} wing test problem } 
\label{sec:naca_Intro}
An analysis of unsteady flow over a \textsc{NACA0012} airfoil is performed in addition to the cylinder test problem. Similar free-stream flow conditions are applied to this test problem, using atmospheric conditions at 5,000 ft. The Mach number is increased to 0.25; however, this flow is still within the subsonic flow regime. The chord length of this symmetric airfoil geometry is set to 0.001 meters, producing flow with a Reynolds number close to 5,000. Experimental analysis suggests that for the \textsc{NACA0012} airfoil, a free-stream flow at this Reynolds number produces predominantly laminar flow over the airfoil and in its wake \cite{ja:huang1995lowRe}, allowing for turbulent effects to be neglected in this analysis. At an angle of attack of 15 degrees, vortices shed into the wake due to laminar separation of the flow over the upper surface of the airfoil generating an unsteady flow solution. As with the cylinder case, Roe's flux scheme is used with no flux limiter.

Figure \ref{fig:naca_flow} shows snapshots of the density field at different time moments. The distribution of Jacobian eigenvalues is shown in Figure \ref{fig:naca_eigs}, and indicates that a large number of fast eigenvalues are clustered together. The performance diagrams in Figure \ref{fig:naca3_M4_steps_timing} indicate that the fastest methods here are Rosenbrock and ROW methods, until ROK and SDIRK become most efficient for errors below $10^{-4}$. This test problem converges to a steady state solution, and this particular dynamics favors the implicit Rosenbrock methods, since the fast modes need  only to be only damped out and not to be accurately solved. 


\begin{figure}
    \centering
\includegraphics[width = \textwidth]{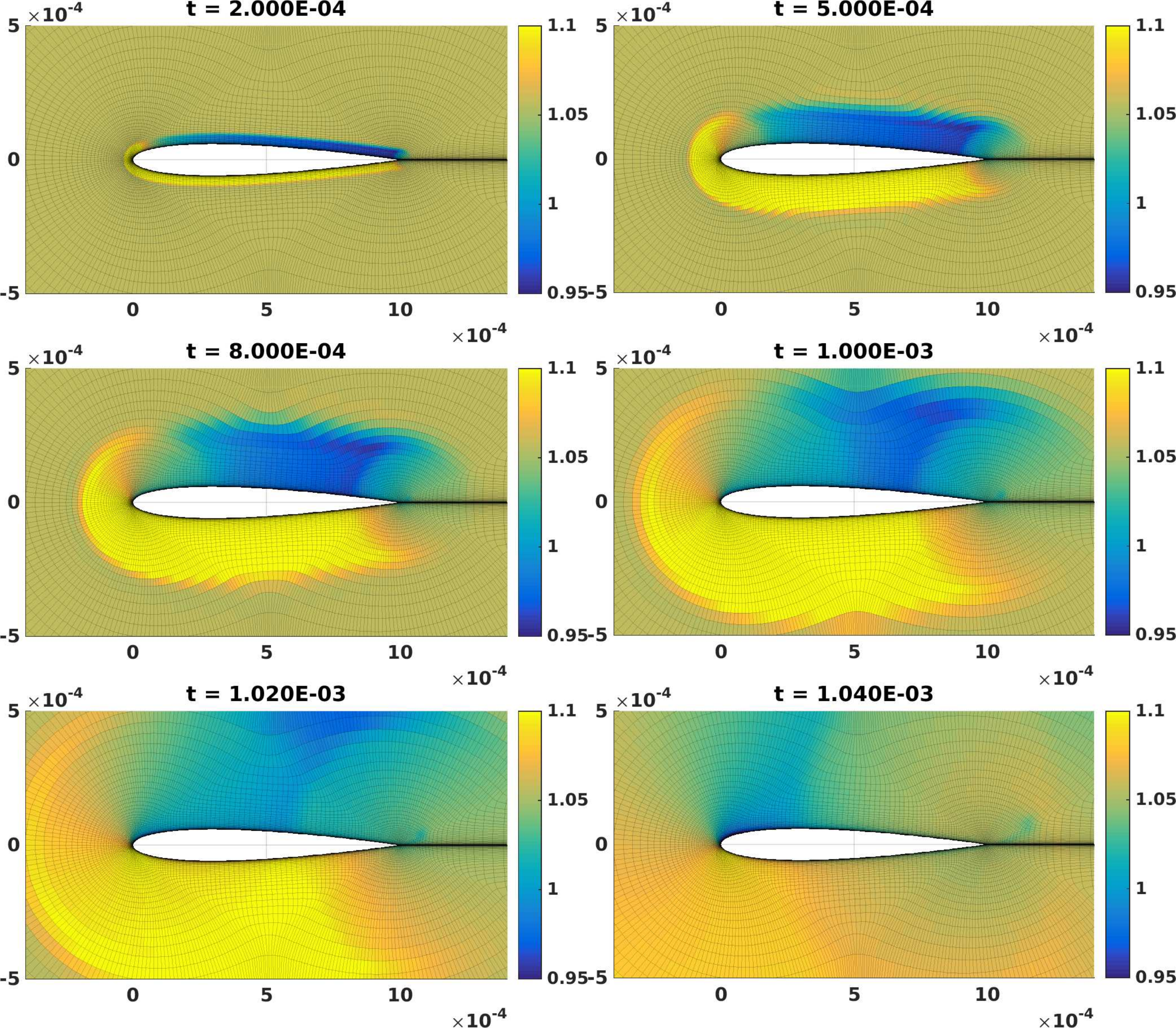}
\caption{Snapshots of density component of the flow for \textsc{NACA0012} at different times.}
\label{fig:naca_flow}
\end{figure}
%
\begin{figure}[htbp]
\centering
 \includegraphics[width=0.7 \linewidth]{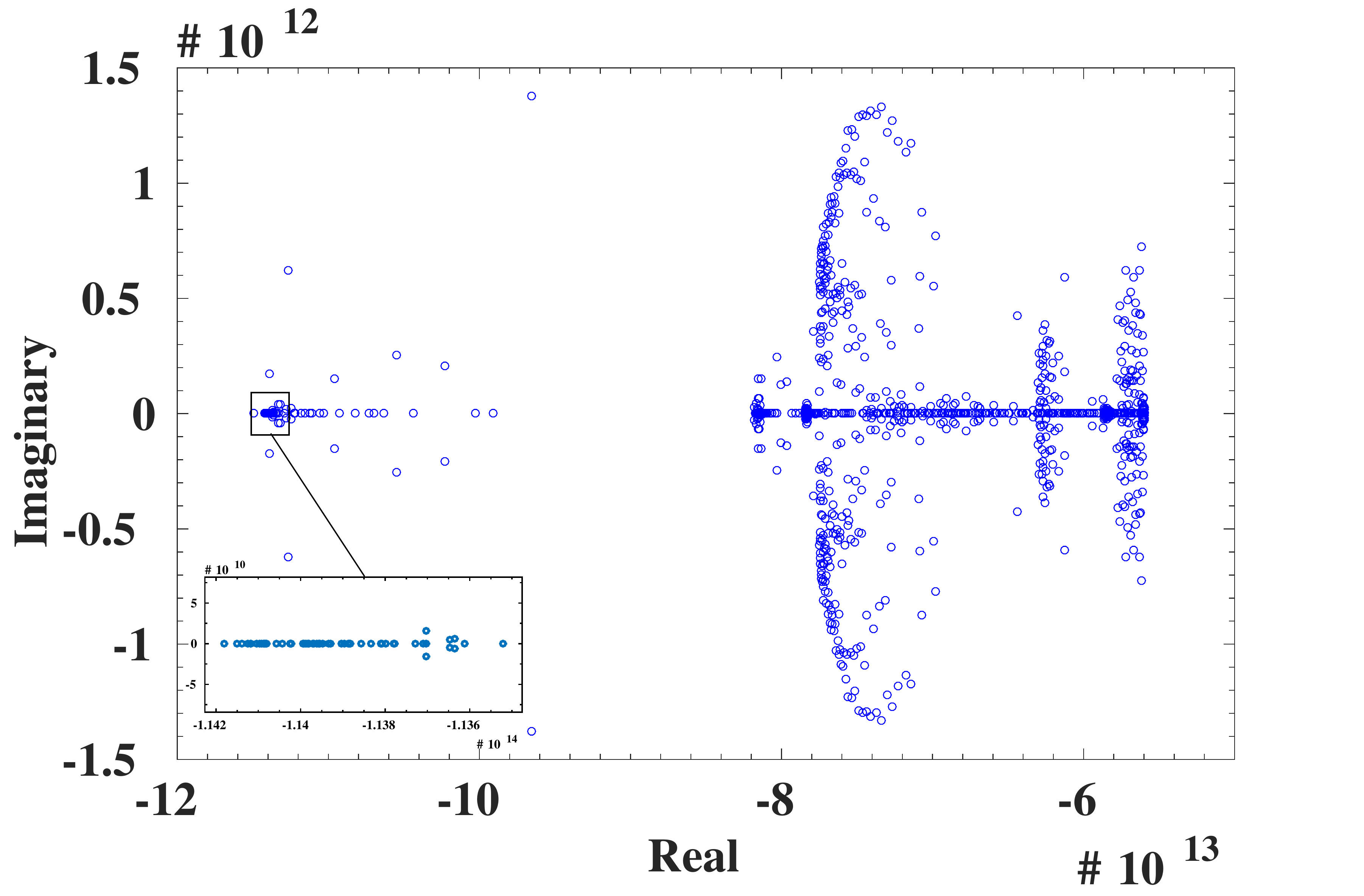}
\caption{Eigenvalue distribution of the Jacobian for the \textsc{NACA0012} test problem.}
\label{fig:naca_eigs}
\end{figure}
\begin{figure}[htbp]
\centering
\begin{subfigure}{.45\textwidth}
  \centering
  \includegraphics[width=\linewidth]{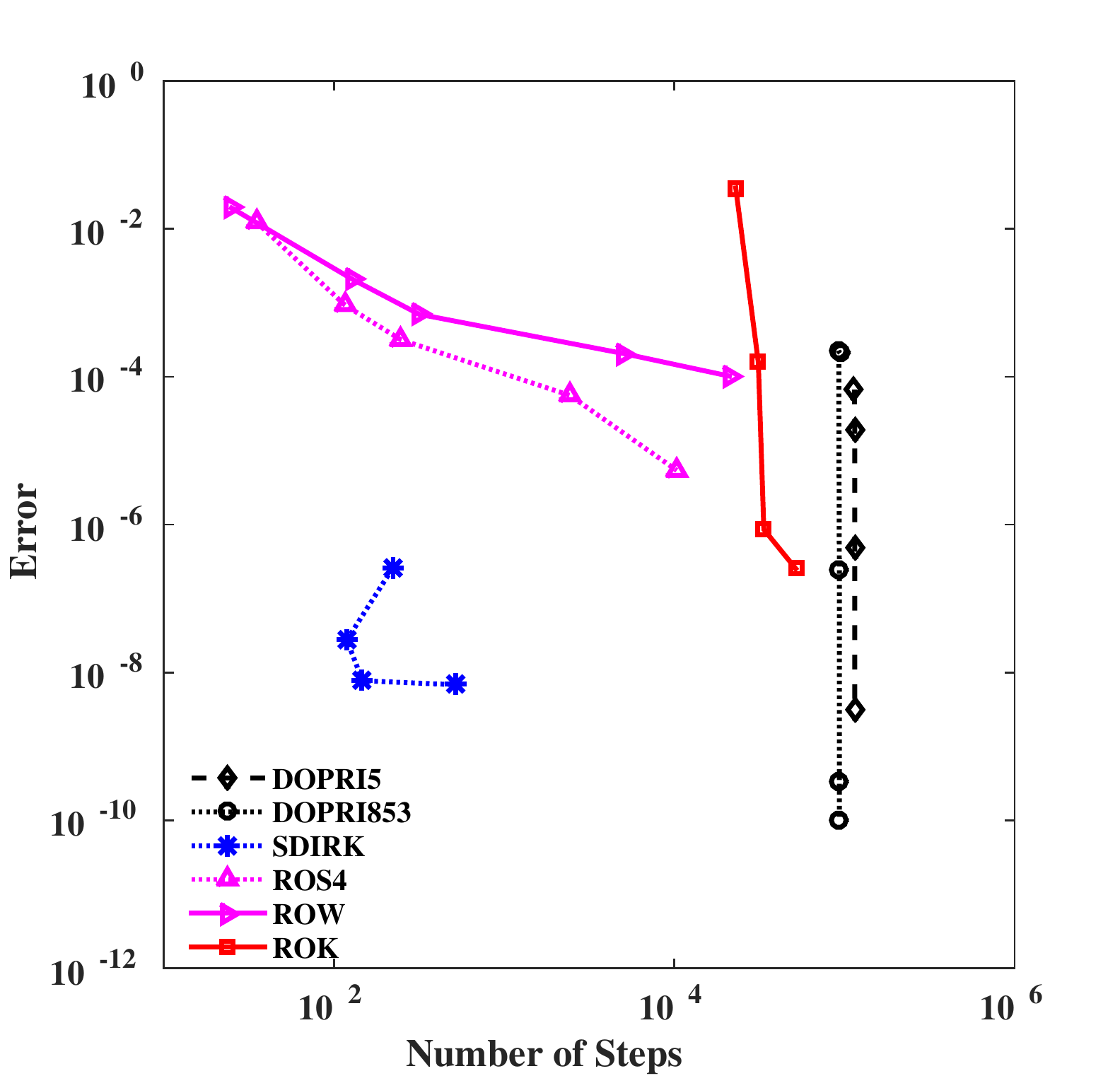}
  \caption{Convergence diagram}
  \label{fig:naca3_M4_steps}
\end{subfigure}%
\begin{subfigure}{.45\textwidth}
  \centering
  \includegraphics[width= \linewidth]{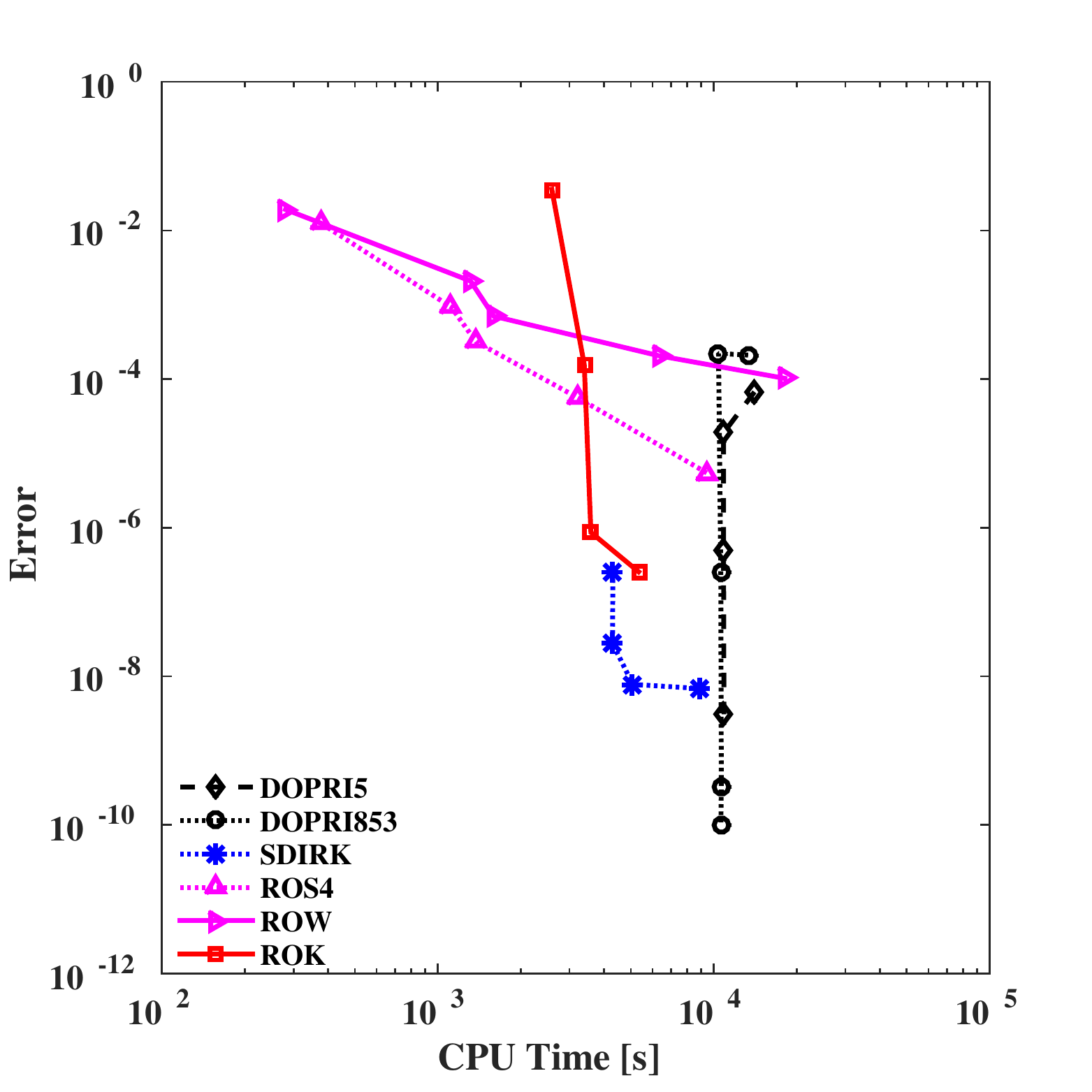}
  \caption{Work-precision diagram}
  \label{fig:naca3_M4_timing}
\end{subfigure}
\caption{Relative performance of different integration methods applied to the \textsc{NACA0012} test problem.}
\label{fig:naca3_M4_steps_timing}
\end{figure}

\subsection{Vortex shedding cylinder with iso-thermal conditions test problem} 

The final test problem uses the vortex shedding cylinder problem 2, with iso-thermal conditions at cylinder boundaries for a wall temperature of $33^{\circ} K$. Among each class of explicit and implicit integrators we have selected the fastest representative methods. We have also chosen three variations of ROK method with 4, 8, and 12 Krylov basis vectors, respectively. The results in Figure \ref{fig:cool_ring_steps_timing} demonstrate that Krylov methods retain their computational superiority for a wide range of error tolerances. Furthermore, we observe that adding more basis vectors to the Krylov subspace requires extra cost (Figure \ref{fig:cooled_ring_timing}) and in turn increases step sizes slightly (Figure \ref{fig:cooled_ring_steps}), but ultimately the most efficient method is the one with minimum basis size, i.e. 4 vectors. Finally, it is worthwhile to point out that once the adaptive error controller of the step size reaches the GMRES tolerance, the error of JFNK methods does not decrease any further. This is seen in Figure \ref{fig:cool_ring_steps_timing} forthe  Rosenbrock-W method where the error curve  flattens at an error level of about 5E-4.

\begin{figure}[htbp]
\centering
\begin{subfigure}{.45\textwidth}
  \centering
  \includegraphics[width=\linewidth]{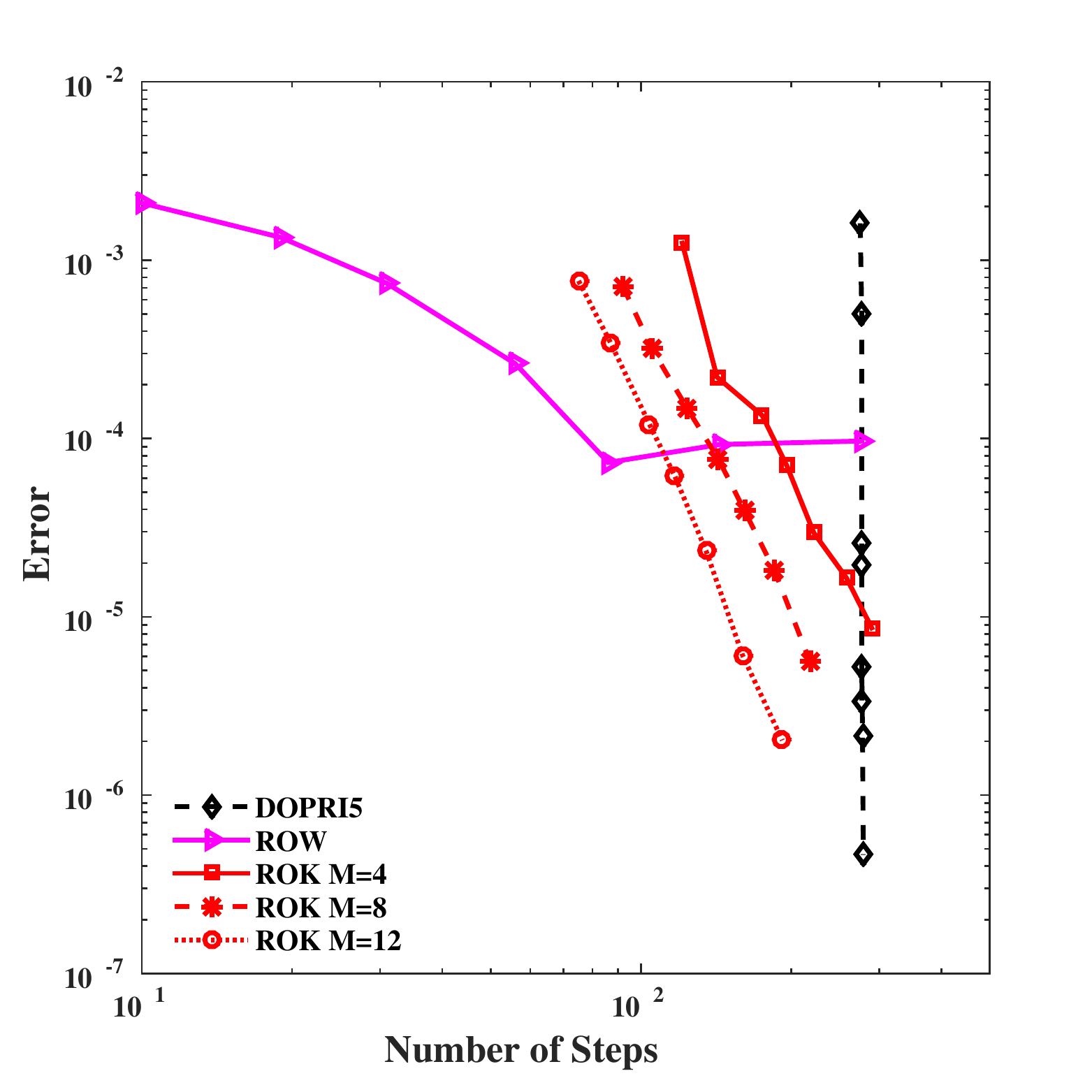}
  \caption{Convergence diagram}
  \label{fig:cooled_ring_steps}
\end{subfigure}%
\begin{subfigure}{.45\textwidth}
  \centering
  \includegraphics[width= \linewidth]{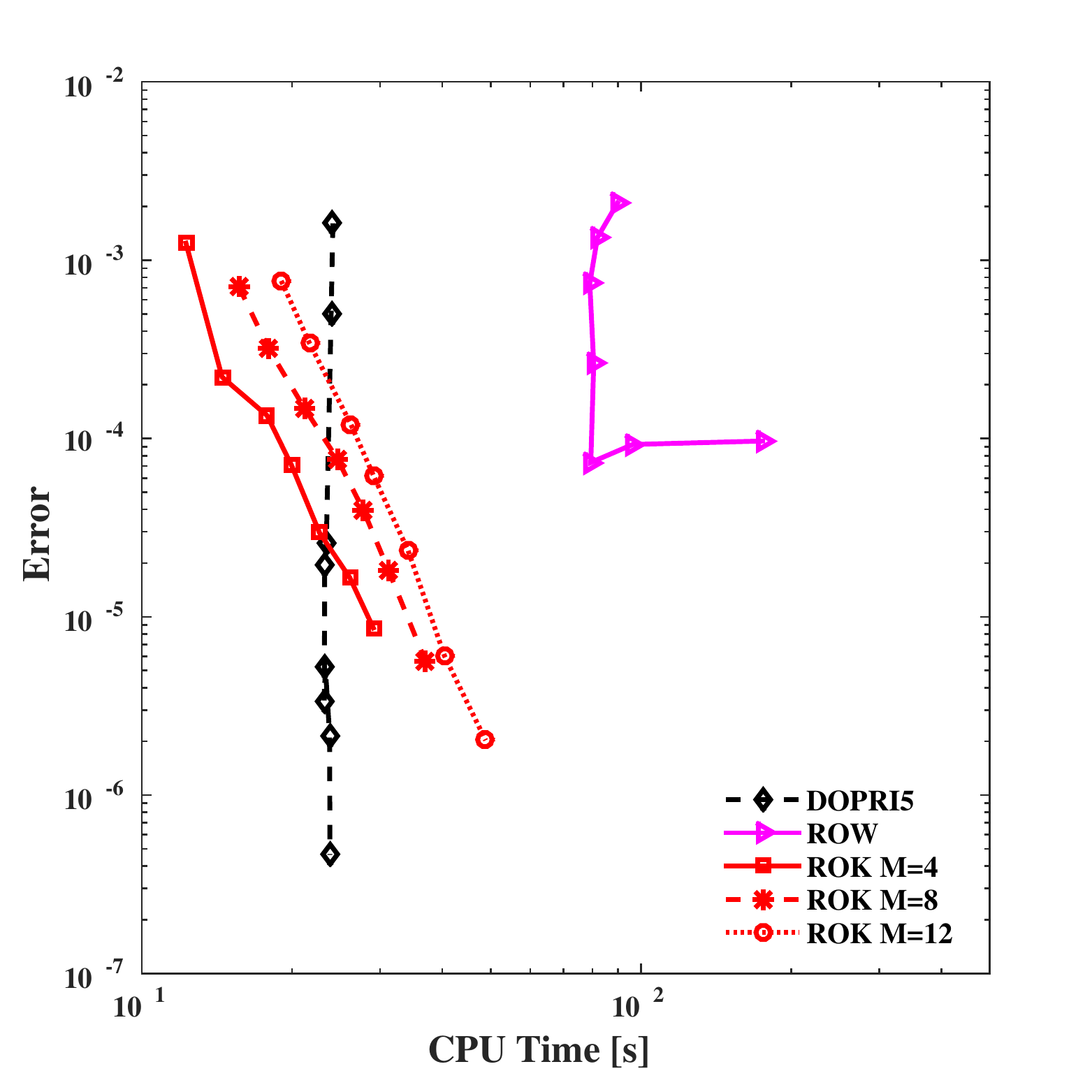}
  \caption{Work-precision diagram}
  \label{fig:cooled_ring_timing}
\end{subfigure}
\caption{Relative performance of different integration methods applied to  cooled cylinder test problem with $T_\textrm{wall} = 33^{\circ} K$.}
\label{fig:cool_ring_steps_timing}
\end{figure}

\section{Conclusions}
\label{sec:Conclusions}

This paper studies the performance of several types of high-order matrix-free time stepping methods when applied to solve unsteady flow problems. The methods under consideration are explicit Runge-Kutta, diagonally implicit Runge-Kutta and Rosenbrock schemes paired with iterative linear algebra solvers, and Rosenbrock-Krylov schemes. All implementations of implicit methods are matrix-free where the necessary Jacobian-vector products are approximated by finite differences; this is a typical setting for solving large-scale CFD applications.

Favorable properties of explicit methods include their easier implementation overhead and the low computational cost per step. As expected, our numerical experiments show that the overall performance of explicit methods deteriorates quickly in the presence of mild stiffness, and for such problems they are not competitive with implicit methods. 

Traditional implicit methods such as SDIRK and Rosenbrock can take large steps on stiff problems, as expected. However, their overall performance depends on how well the underlying linear systems are solved at each step. We observed that inaccurate linear solutions lead to loss of convergence, and that in absence of well-tuned linear algebra preconditioners the computational costs of matrix-free implicit methods are quite large. 

Rosenbrock-Krylov methods are the most effective when applied to flow problems with a limited number of stiff modes. These methods are a suitable choice when exact full Jacobians are not available, as is the case in large CFD problems. The order conditions theory of Krylov-based methods accounts for the errors associated with linear algebra; indeed, our numerical experiments confirm that they show full order where other implicit methods suffer from reduced temporal convergence caused by  inexact Jacobian approximations or by poorly converged linear system solutions. On the other hand, as the test problems become more  stiff,  Krylov-based methods lose their performance superiority due to the increased computational cost of creating a large Krylov basis required for numerical stability.

The numerical experiments in this study were performed on a 2D Navier-Stokes problem. Extension of the investigation to 3D unsteady flow problems is a future direction to pursue. Furthermore, finding automatic strategies for the cost-effective construction of Krylov basis is an important question for the advancement of Krylov-based family of methods.

\section{Acknowledgments}
This work was funded by the Air Force Office of Scientific Research (AFOSR) computational mathematics program via grant No. FA9550-12-1-0442 and by Computational Science Laboratory (CSL) in the Department of Computer Science at Virginia Tech.
\section{Appendix: Order reduction with matrix-free methods }
\label{sec:apex}
In this section we confirm the orders of convergence for the time-stepping methods used in this paper by applying them to Lorenz--96 test problem \cite{lorenz1996predictability}. Proposed by Edward Lorenz, this model is often used to model chaotic behavior of atmospheric systems:
\begin{align*}
\frac{\partial u_i} {\partial t} = \left( u_{i+1} - u_{i-2} \right) \, u_{i-1} - u_i + F \qquad \text{for} \quad i=1, \cdots, 40.
\end{align*}
The Jacobian of this system is a banded matrix that can be implemented in sparse format and used in the implicit methods of Section \ref{sec:Time_Integration_CFD}. Setting the external force factor $F=8$ and using a range of fixed time steps to propagate the model forward, we observe nearly full order for all of the methods as reported in Table \ref{tab:Lorenz--96_conv}. One source of local error causing order reduction can be traced back to the truncation errors made by replacing the Jacobian-vector product in equation \eqref{eqn:matvecFD} with a first-order finite difference approximation. Poor convergence of the Newton's iteration as well as the Krylov-based solver for the linear system in the implicit methods are other sources of error contributing to inexact stage vectors and ultimately to order reduction. It is notable, however, that in all the convergence tests the Rosenbrock-Krylov method retains its full order of convergence. Unlike fully and linearly implicit methods, the system solved in equation \eqref{eqn:ROK_Lambda_system} is formed using the Arnoldi iteration with an inexact Jacobian from the beginning and is solved using a direct method and therefore the issues arising from the convergence of the iterative solvers is avoided altogether. Interested readers may consult \cite{Sandu_2014_ROK} for a detailed analysis of this phenomenon.

\begin{table}[h]
\centering
\caption{Orders of convergence for methods applied to the Lorenz--96 problem.}
\label{tab:Lorenz--96_conv}
\begin{tabular}{lcc}
\hline
Method          			& Numerical order      & Theoretical order\\
\hline
ERK4            			& 4.00				& 4 \\
ERK5 (DOPRI5)   			& 5.58 				& 5 \\
ERK5 (DOPRI853) 			& 6.14 				& 8 \\
SDIRK           			& 3.89				& 4 \\
ROS4            			& 3.91  			& 4 \\
ROW						    & 2.94 				& 3 \\
ROK             			& 3.85 				& 4\\
ODE45           			& 5.47 				& 5 \\
\hline
\end{tabular}
\end{table}
%
\begin{figure}[htbp]
\centering
  \includegraphics[width=0.5 \linewidth]{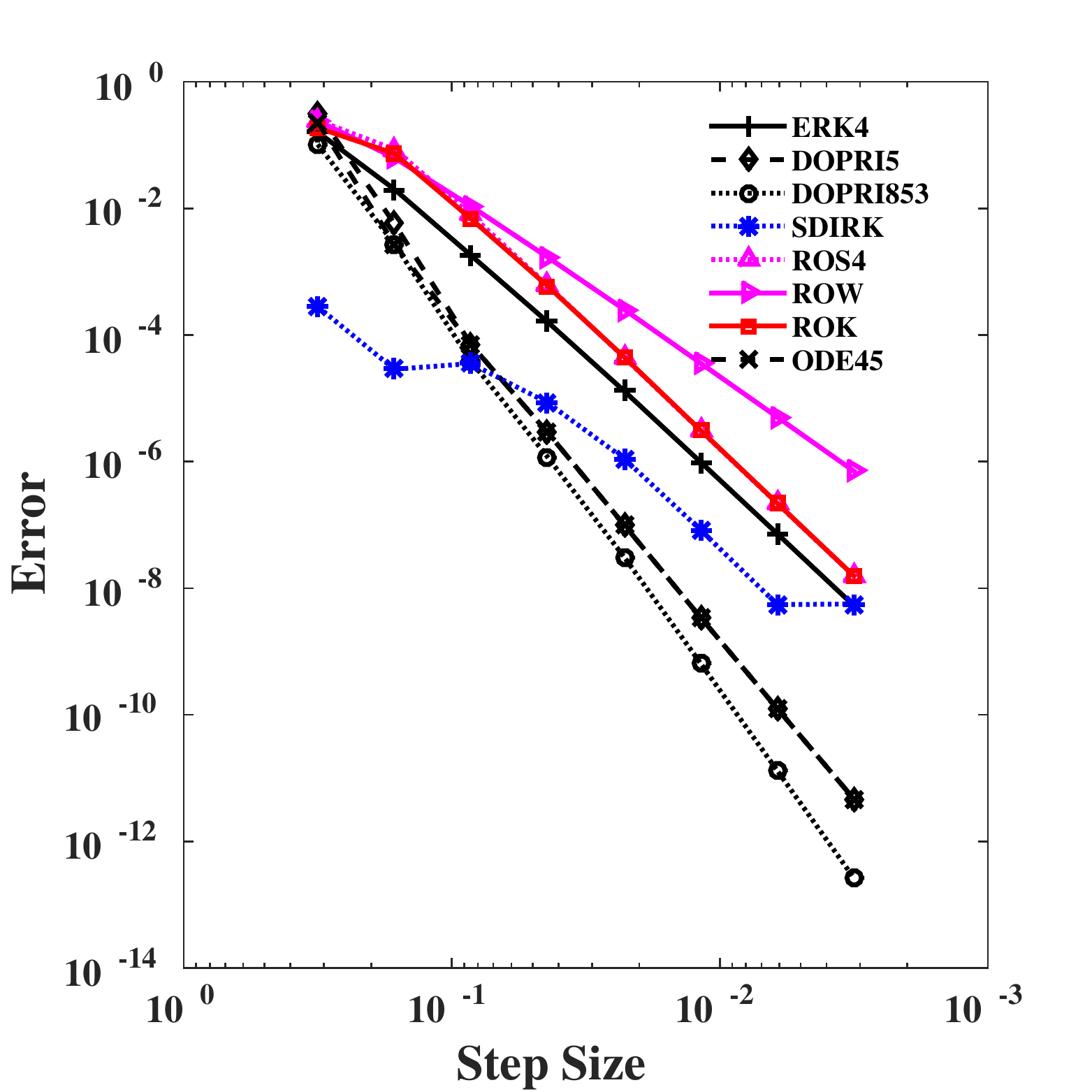}
\caption{Convergence diagram for the Lorenz--96 test problem.}
  \label{fig:Conv_Lorenz}
\end{figure}

\section*{References}

\bibliography{mybibfile,ode_general,pde_time_implicit,pde_time_explicit,ode_rosenbrock,sandu}

\end{document}

%% file: logo.tex
\thispagestyle{empty}
\setcounter{page}{0}

\makeatletter
\def\Year#1{%
  \def\yy@##1##2##3##4;{##3##4}%
  \expandafter\yy@#1;
}
\makeatother

\begin{Large}
\begin{center}
Computational Science Laboratory Technical Report CSL-TR-\Year{\the\year}-{6} \\
July 26, 2016
\end{center}
\end{Large}
\vfil

\vfil
\begin{huge}
\begin{it}
\begin{center}
``{\tt A Numerical Investigation of Matrix-Free Implicit Time-Stepping Methods for Large CFD Simulations}''
\end{center}
\end{it}
\end{huge}
\vfil

\vfil
\begin{large}
	\noindent
			\textbf{Cite as}:\tt{
			A Sarshar, P Tranquilli, B Pickering, A McCall, C J Roy, A Sandu, ``A numerical investigation of matrix-free implicit time -stepping methods for large CFD simulations'', Computers \& Fluids, Vol. 159, 15 Dec. 2017, PP. 53-63.\\ \url{https://doi.org/10.1016/j.compfluid.2017.09.014}}
\end{large}
\vfil

\begin{large}
\begin{center}
Computational Science Laboratory \\
Computer Science Department \\
Virginia Polytechnic Institute and State University \\
Blacksburg, VA 24060 \\
Phone: (540)-231-2193 \\
Fax: (540)-231-6075 \\ 
Email: \url{sarshar@vt.edu} \\
Web: \url{http://csl.cs.vt.edu}
\end{center}
\end{large}

\vspace*{1cm}

\begin{tabular}{ccc}
\includegraphics[width=2.5in]{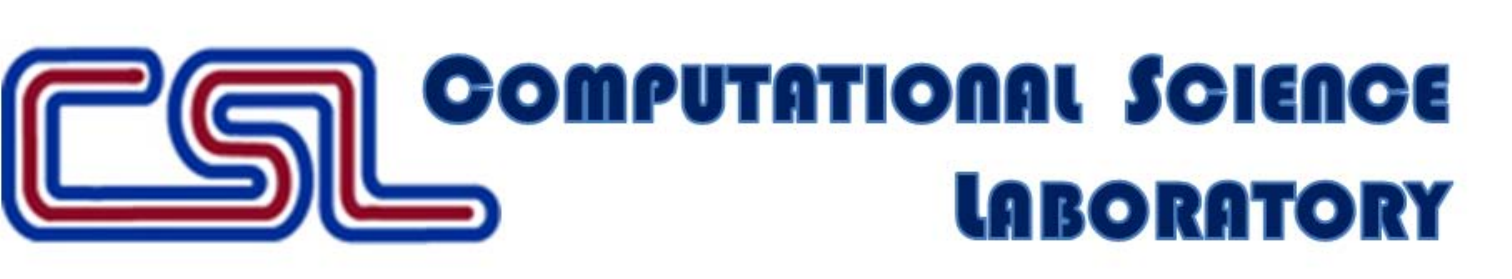}
&\hspace{2.5in}&
\includegraphics[width=2.5in]{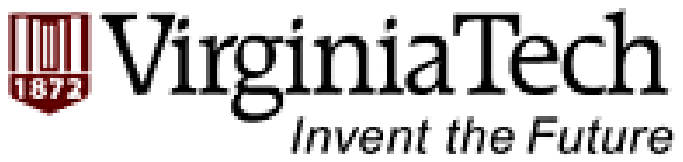} \\
{\bf\em\large Compute the Future} &&\\
\end{tabular}

\newpage